\documentclass[useAMS,usenatbib]{mn2e}
\usepackage{graphicx}
\usepackage{epsfig}
\usepackage{multirow}
\usepackage{float}
\usepackage{color}
\usepackage{enumerate}
\usepackage{verbatim}  
\usepackage{booktabs}
\usepackage{amsmath}
\usepackage[table]{xcolor}
\usepackage[section]{placeins}
\usepackage{bm}

\title[Understanding the residual patterns of radio pulsars] {Understanding the residual patterns of timing solutions of radio pulsars with a model of magnetic field oscillation}

\author[X. D. Gao et al.]
{Xu-Dong Gao$^{1,2}$ \thanks{E-mail: xdgao@mail.bnu.edu.cn}, Shuang-Nan Zhang$^{2,3}$ \thanks{E-mail: zhangsn@ihep.ac.cn}, Shu-Xu Yi$^{3,4}$, Yi Xie$^{2,4}$, and Jian-Ning Fu$^{1}$\\
$^{1}$ Department of Astronomy, Beijing Normal University, Beijing 100875, China\\
$^{2}$ National Astronomical Observatories, Chinese Academy Of Sciences, Beijing 100012, China\\
$^{3}$ Key Laboratory of Particle Astrophysics, Institute of High Energy Physics, Chinese Academy of Sciences, Beijing 100049, China\\
$^{4}$ University of Chinese Academy of Sciences, Beijing 100049, China
}

\begin{document}

\date{}

\maketitle

\label{firstpage}

\begin{abstract}
We explain some phenomena existing generally in the timing residuals: amplitude and sign of the second derivative of a pulsar's spin-frequency ($\ddot\nu$), some sophisticated residual patterns, which also change with the time span of data segments. The sample is taken from Hobbs et al.\,(2010), in which the pulsar's spin-frequency and its first derivative have been subtracted from the timing solution fitting. We first classify the timing residual patterns into different types based on the sign of $\ddot\nu$. Then we use the magnetic field oscillation model developed in our group \citep{zhang12a} to fit successfully  the different kinds of timing residuals with the Markov Chain Monte Carlo method. Finally, we simulate the spin evolution over 20 years for a pulsar with typical parameters and analyze the data with the conventional timing solution fitting. By choosing different segments of the simulated data, we find that most of the observed residual patterns can be reproduced successfully. This is the first time that the observed residual patterns are fitted by a model and reproduced by simulations with very few parameters. From the distribution of the different residual patterns in the $P-\dot P$ diagram, we argue that (1) a single magnetic field oscillation mode exists commonly in all pulsars throughout their lifetimes; (2) there may be a transition period over the lifetimes of pulsars, in which multiple magnetic field oscillation modes exist.

\end{abstract}


\section{Introduction}
\label{sec-introduction}

Radio pulsars are renowned for unique high-precision clocks in the universe driven by their stable rotation. The pulsar timing method provides a model of the pulsar's astrometric, orbital and rotational parameters to compare with the observed pulse times of arrival (TOAs) \citep{manchester77,lyne04,lorimer05,edwards06}. Timing residuals of a pulsar are defined as the differences between the predicted TOAs and the actual TOAs. However, radio pulsars exhibit two main timing irregularities, namely `glitches' that are sudden increases in spin rate followed by a period of relaxation, and `timing noise' that shows long-term and stochastic deviations from a regular spin-down model and mainly consists of low-frequency structures.

Timing noise can be characterized as a random walk in the rotational phase, angular velocity or torque and describes the un-modelled red noise in the observed TOAs \citep{cordes85,shannon10}, implying a process autocorrelated on a time-scale of hours to years \citep{cordes80}. Previously the analyses of timing noise have been limited by the relatively short data spans. Only a few papers have analysed a small number of pulsars for long data spans \citep{baykal99,shabanova95,stairs00,shabanova01}. Most of the analyses are mainly about obtaining high-quality power spectral estimates of the timing residuals or fitting the timing residuals of a single pulsar with a simple model. Recently Hobbs et al.\,(2010) (hereafter H10) carried out so far the most comprehensive survey of timing irregularities in 366 pulsars over time-scales longer than a decade, which have characteristic ages larger than $10^{4}$ years. Based on the simple spin-down model, the second frequency derivative can be modelled by the cubic terms of a Taylor series as follows,
\begin{equation}\label{Phi0}
\Phi(t)= \Phi_{0}+\nu(t-t_{0})+\frac{1}{2}\dot\nu(t-t_{0})^{2}+\frac{1}{6}\ddot\nu(t-t_{0})^{3}+...,
\end{equation}
where $\Phi_{0}$ is the phase at time $t_{0}$. However, values of $\ddot\nu$ obtained from timing fits are typically orders of magnitude larger than the prediction of the vacuum dipole model and usually have different signs. In the meantime, the braking indices determined by these derivatives have anomalous values range from $-2.6\times10^{8}$ to $+2.5\times10^{8}$. Therefore, Hobbs et al. argued that these cubic terms are not dominated by the intrinsic dipole braking of pulsars and only fit the pulse frequency and its first derivative and consider the remaining features as timing noise. Subsequently, They tried to categorize the pulsars based on the structures existing in their timing residuals. Nevertheless, they argued that two problems exist in their categorization, namely, the TOA precision achievable and the data spans available.

However, the physical processes behind timing noise have not been well explained. The phenomenon is attributed to various mechanisms. It has been suggested that the timing noise is dominated by the recovery from unseen glitches \citep{johnston99}. As the pulsar ages, glitch activity decreases and the timing noise is dominated by changes in the magnetosphere. Lyne et al.\,(2010) demonstrated that the observed timing noise correlates with the observed pulse shape, which indicates that some timing noise may be caused by magnetospheric processes. Some authors argued that timing noise may be driven by variability in the coupling between the crust and superfluid interior \citep{alpar86,jones90} and fluctuations in the external spin-down torque \citep{cheng87a,cheng87b,urama06}. A more complete understanding of timing noise will provide us an insight into the interior structure of neutron stars.

An additional possibility is that the magnetic fields of pulsars decay with age. Baym et al. (1969) gave the first estimate of the characteristic decay time of their magnetic fields. Sang $\&$ Chanmugam (1987) argued that the field does not decay exponentially during the lifetime of the neutron stars. However, a complete theoretical model that could explain all observations does not exist yet, as the evolution of the magnetic field depend on its configuration that is still unknown and even the origin of the field is not clear until now. In general, the magnetic field is often assumed to occupy a substantial fraction of a neutron star's volume and pass through its core or is confined to relatively not very deep layers. Recently, three physical processes were proposed to explain the evolution of the magnetic field inside a neutron star: (1) Hall drift, which is the advection of the magnetic field because of the motion of the free electrons; (2) Ohmic dissipation, which converts magnetic energy to heat because of the finite conductivity of the crust; (3) ambipolar diffusion, which is the interaction of the electric currents with the neutrons deeper inside a neutron star \citep{goldreich92}. On the other hand, population synthesis studies suggested that old pulsars show no significant magnetic decay over their lifetimes \citep{regimbau01,faucher06}, although the opposite conclusion has also been claimed \citep{gonthier04,popov10}. These contradictions may be resolved by the assumption that the neutron star magnetic field is maintained by two current systems. Long living currents in the superconducting core support the large-scale dipolar field and are responsible for the spin down of old pulsars, but currents in the crust support the short-lived part of the field \citep{pons07}.

The estimates of the magnetic field strength of neutron stars usually come from radio pulsars with measured spin-down rates. As the real ages of these pulsars are usually unknown, one can determine their spin-down age, $\tau_{c}=P/2\dot P$ , where $P$ is the spin period of the pulsar, as the indicators of the true ages of pulsars. However, Zhang $\&$ Xie (2011) showed that spin-down ages are normally significantly larger than the ages of the supernova remnants physically associated with them, which in principle should be the unbiased age indicators of the pulsars \citep{lyne75,geppert99}. It may be evidence of the magnetic field decay over their lifetimes, as the decay can alter the spin-down rate of a pulsar significantly. To explain the observed periodic or quasi-periodic evolution of the spin of pulsars, it is natural to introduce some oscillation parameters. Zhang $\&$ Xie (2012a) proposed a magnetic field evolution model, which consists a long-term decay modulated by short-term oscillations and can explain the observed statistical properties of $\ddot\nu$ well.

In this paper, we will mainly explain the reason for amplitude and sign of second frequency derivative and the
structure of timing residuals of a given pulsar varying with the time span of data segments. By contrasting with Hobbs's opinion, we argue that timing residuals, which have subtracted pulsar's spin frequency and its first derivative and are dominated by $\ddot\nu$ (i.e. Fig.\,3 in H10), are still due to the evolution of magnetic field of pulsars. The structure of the paper is as follows. In the next section, we introduce the sample selection of the pulsars, and classify the timing residuals based on the the sign of $\ddot\nu$ and the detailed structure of timing residuals. In section 3, we review magnetic field evolution model we developed previously and use Markov Chain Monte Carlo (MCMC) method to fit the timing residuals with a single oscillation component and two oscillation components of the model. Then we reclassify the timing residuals based on the number of oscillation components. In section 4, we discuss the evolution of pulsars through their distribution of different kinds of timing residuals in the $P-\dot P$ diagram and physical implication of the oscillation term. The summary of our results is given in section 5.

\section{Classify the Timing Residuals}
\label{sec-classify-residuals}

As we mentioned above, Hobbs et al.\,(2010) attempted to categorize pulsars based on the structures existing in their timing residuals: (1) 37\% of the timing residuals are dominated by the measurement errors and show no features; (2) 20\% have residuals that can be modelled by a significant positive $\ddot\nu$ value; (3) 16\% have cubic terms that correspond to a negative $\ddot\nu$ value; and (4) 27\% show more complicated structures. In the meantime, they showed that their simple categorization has two problems: (1) The TOAs of some pulsars are measured more precisely than the TOAs of others (see Fig.\,4 in H10); (2) The structure of timing residuals of any pulsar is likely to change as the data span varies (see Fig.\,5 in H10).

To better classify the timing residuals, we try to separate the pulsars that their timing residuals have detailed structures and those that have no structures in the residuals from the H10 sample. Figure 1 shows the distributions of $\sigma_{3}$ and $\sigma_{1}$ of all 366 pulsars taken from H10, where $\sigma_{1}$ represents the un-weighted rms of the residuals after fitting for $\nu$ and $\dot\nu$, and $\sigma_{3}$ represents the rms after whitening the data set by fitting and removing harmonically related sinusoids \citep{hobbs04,hobbs10}. From these distributions alone we can not identify those pulsars with obvious structures in their timing residuals. In Figure 2 we plot the distribution of $\sigma_{3}/\sigma_{1}$ of all 366 pulsars we take from H10. We find that there are two distinctly different peaks in Figure 2; the pulsars in one peak that the values of $\sigma_{3}/\sigma_{1}$ are approximately 0 have distinct structures of timing residuals as shown in Figures 3 and 4, but in the other peak no distinct structure exists, which means that the residuals after subtraction of the pulsar's $\nu$ and $\dot\nu$ are approximately white noise. Therefore, the pulsars in our sample are selected with $\sigma_{3}/\sigma_{1} < 0.4$, which have periodic or quasi-periodic structures in their timing residuals; the number of pulsars with $\sigma_{3}/\sigma_{1} > 0.4$ or $\sigma_{3}/\sigma_{1} < 0.4$ are 166 and 200, respectively. In Figure 5 we plot the distributions of $\sigma_{1}$ and $\sigma_{3}$ of the two samples; we can not distinguish the two samples based on $\sigma_{1}$ or $\sigma_{3}$ individually, since their main peaks overlap in the distributions.

\begin{figure}
\centering
\includegraphics[scale=0.4,width=\columnwidth]{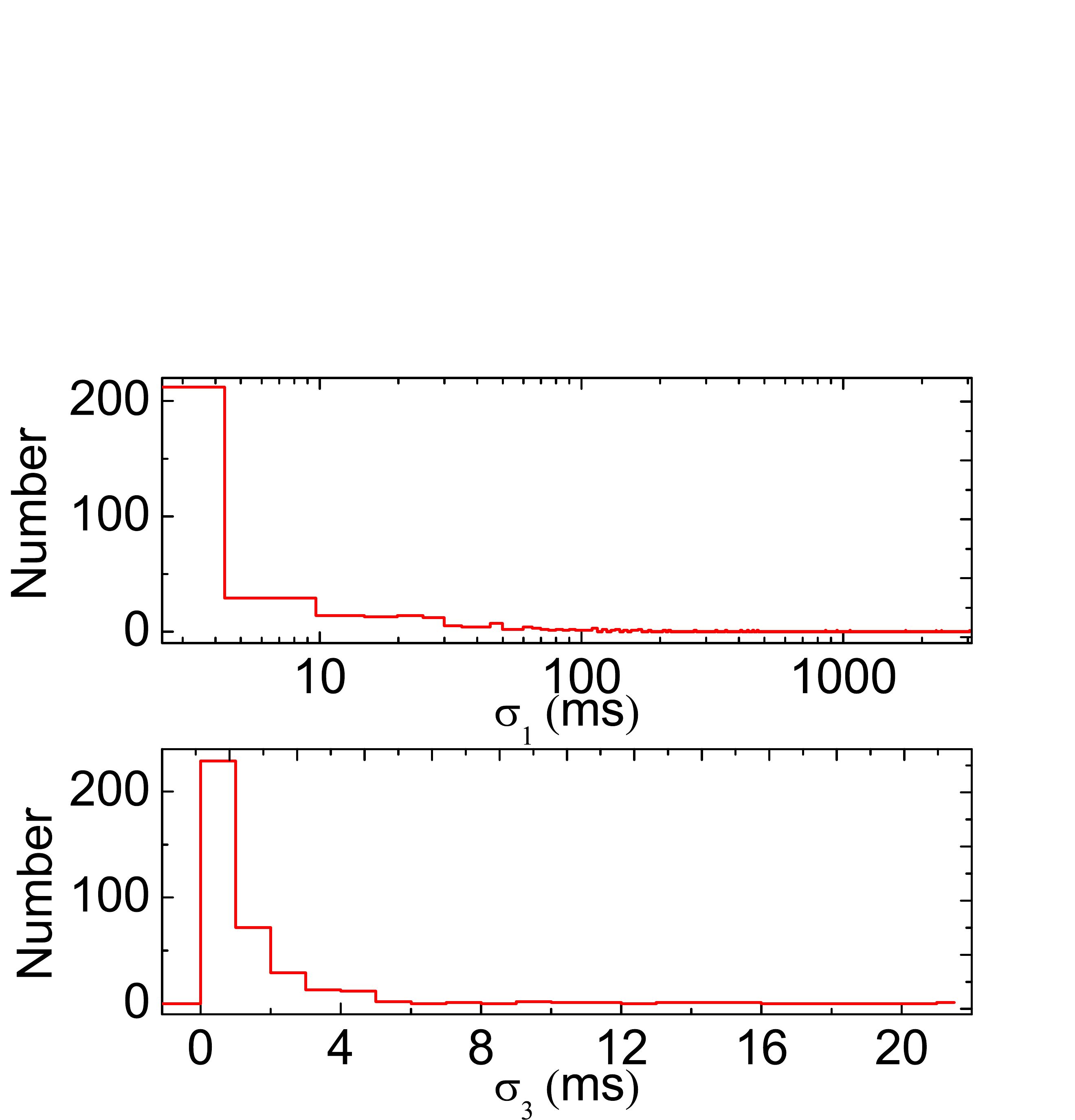}
\caption{Distributions of $\sigma_{1}$ and $\sigma_{3}$ of all 366 pulsars, where $\sigma_{1}$ represents the unweighted rms of the residuals after fitting for $\nu$ and $\dot\nu$, and $\sigma_{3}$ represents the rms after whitening the data set by fitting and removing harmonically related sinusoids.}
\label{dis}
\end{figure}

\begin{figure}
\centering
\includegraphics[scale=0.4,width=\columnwidth]{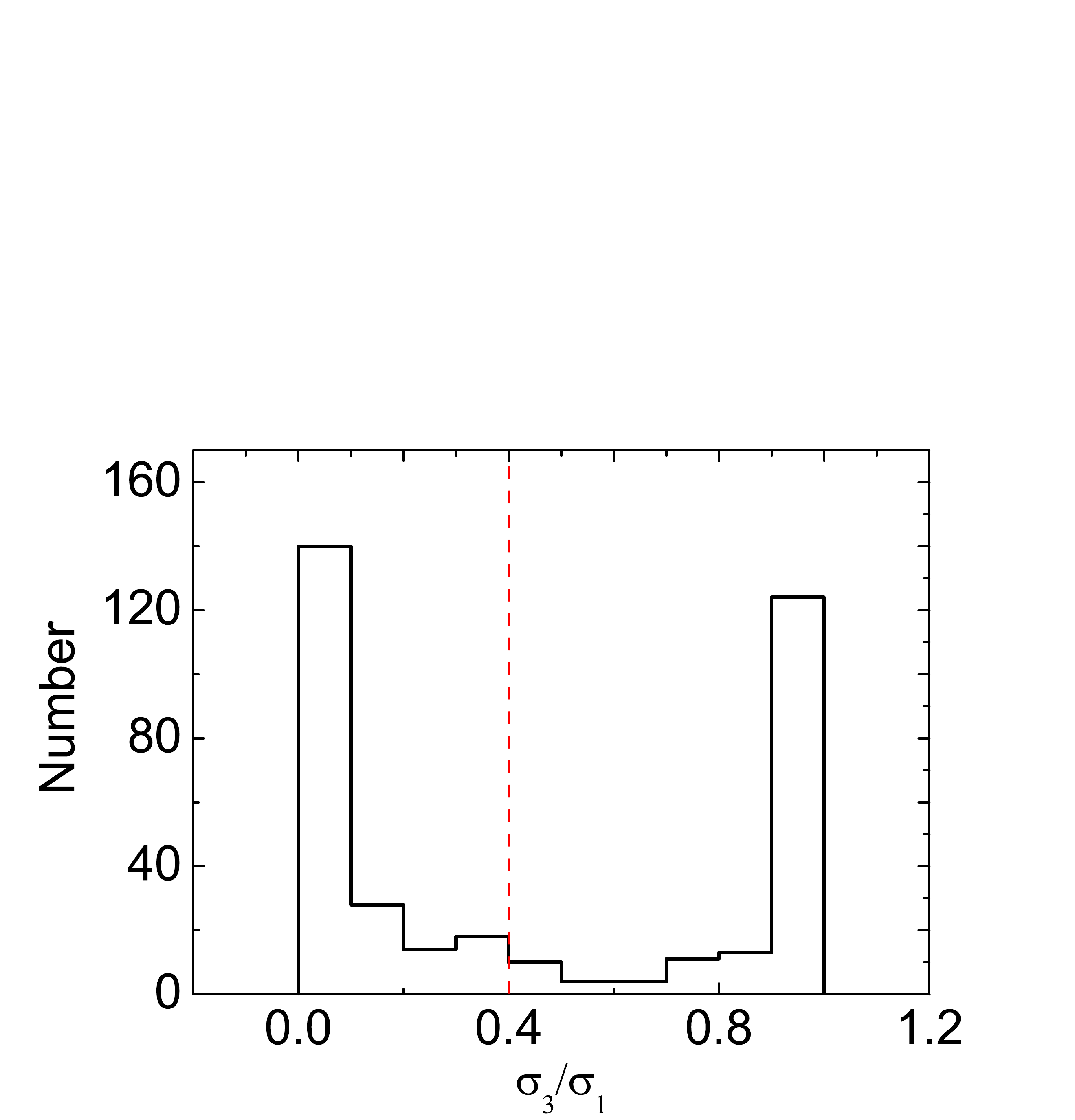}
\caption{Distribution of $\sigma_{3}/\sigma_{1}$ of all pulsars. The number of $\sigma_{3}/\sigma_{1} < 0.4$ or $\sigma_{3}/\sigma_{1} >0.4$ are 200 and 166, respectively.}
\label{dis}
\end{figure}

\begin{figure}
\centering
\includegraphics[angle=0,scale=0.4,width=\columnwidth]{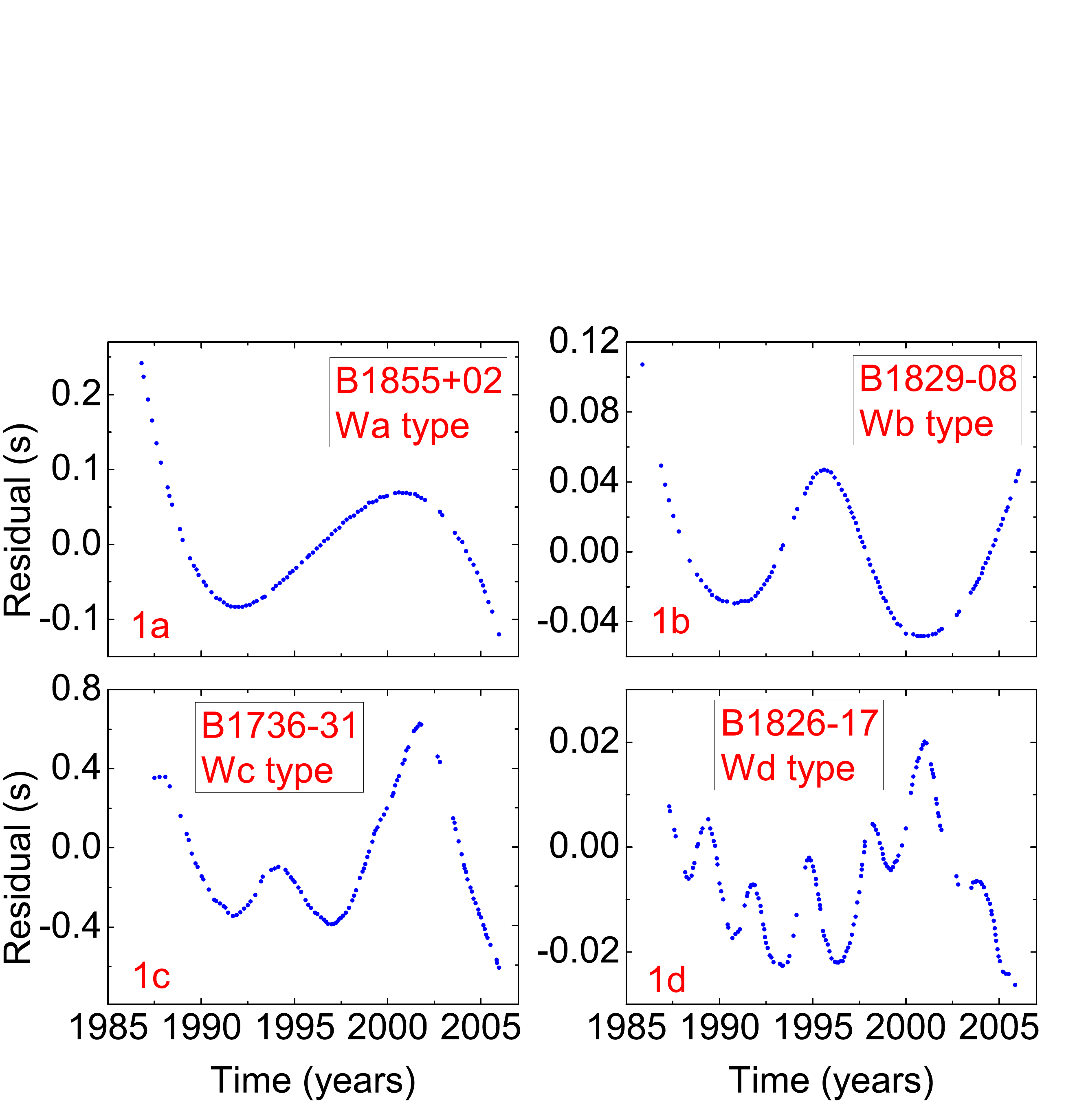}
\caption{$W$ mode timing residuals. Each sub-panel represents the different subclass, for which the pulsar's spin-frequency and its first derivative have been fitted and removed.}
\label{W-mode}
\end{figure}

\begin{figure}
\centering
\includegraphics[angle=0,scale=0.4,width=\columnwidth]{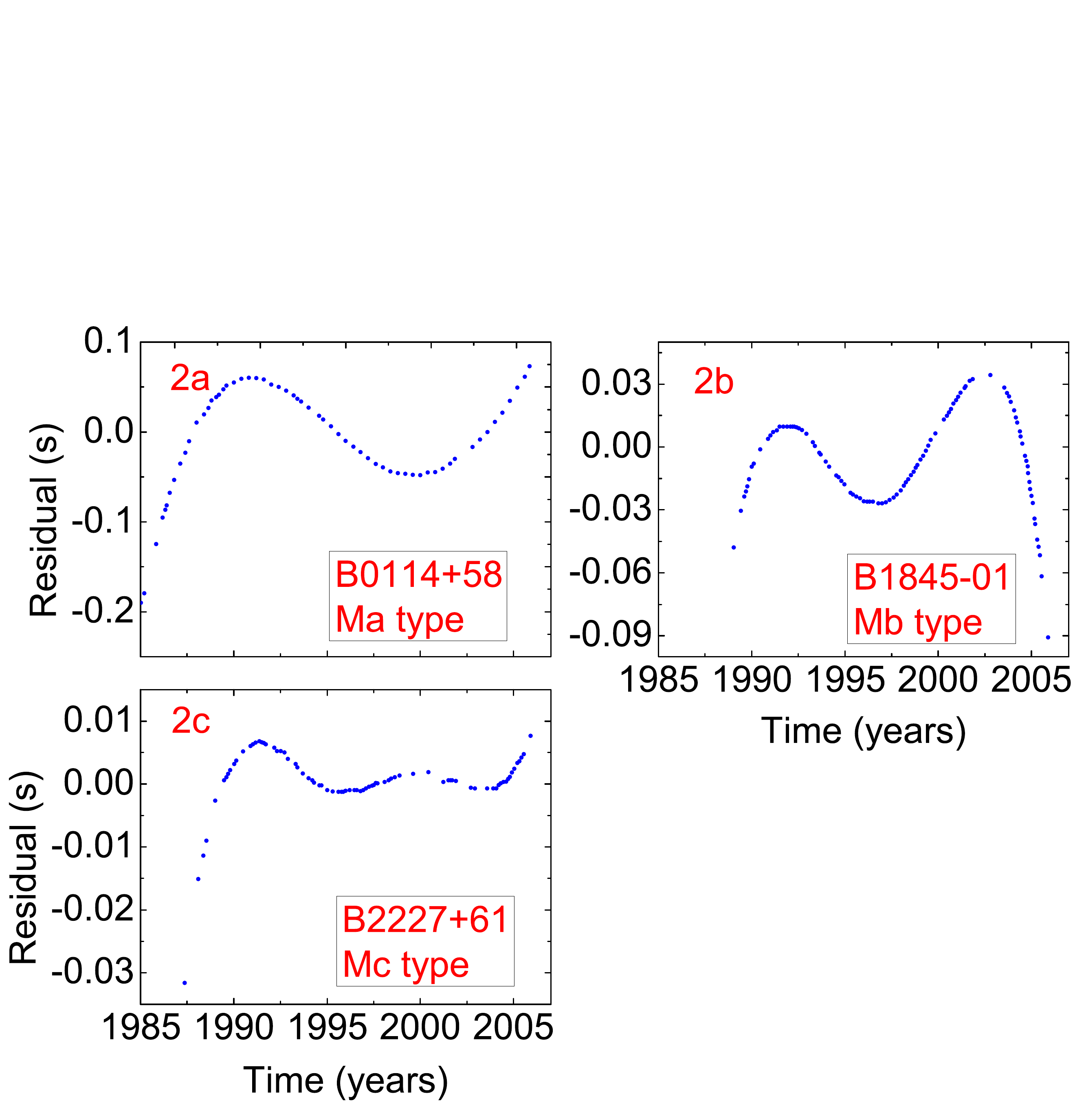}
\caption{$M$ mode timing residuals. Each sub-panel represents the different subclass, for which the pulsar's spin-frequency and its first derivative have been fitted and removed.}
\label{M-mode}
\end{figure}

\begin{figure}
\centering
\includegraphics[scale=0.4,width=\columnwidth]{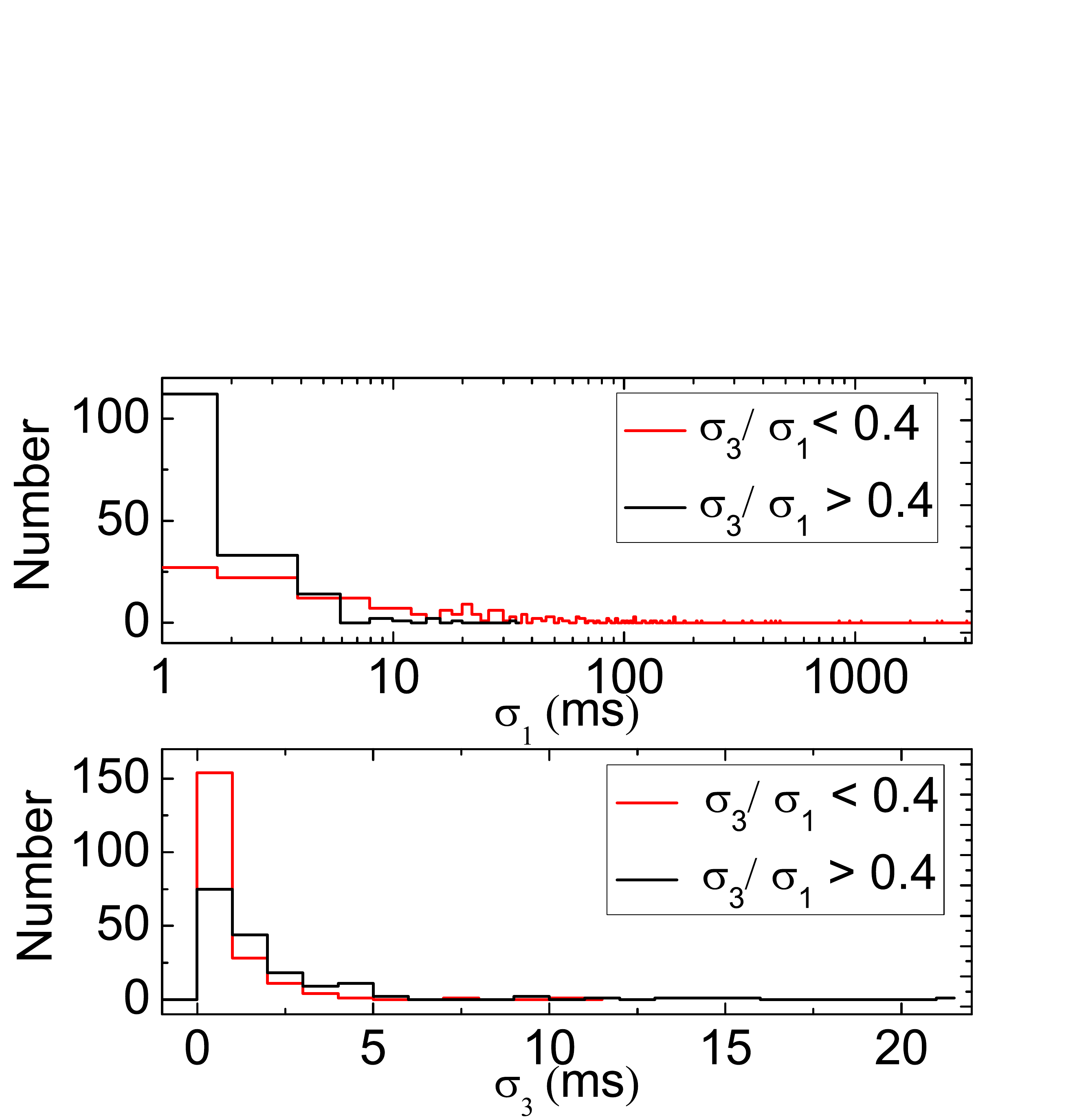}
\caption{Distribution of $\sigma_{1}$ and $\sigma_{3}$ of two samples, respectively.}
\label{dis}
\end{figure}

We thus select the sample that have detailed structures to study the evolution of $\ddot\nu$ with their timing residuals. In the first step, we classify the timing residuals based on the sign of $\ddot\nu$. In addition, based on the structural tendency of timing residuals, we define the residuals of $\ddot\nu >0$ as $W$ mode (see Fig.~\ref{W-mode}). Conversely, we define the residuals of $\ddot\nu <0$ as $M$ mode (see Fig.~\ref{M-mode}). According to equation~(\ref{Phi0}), to the timing residuals after substraction of the pulsar's spin-frequency and its first derivative, the overall structural tendency is dominated by the sign of $\ddot\nu$. That is to say, the overall shapes of these two modes (see Fig.\,3a and Fig.\,4a respectively) are the basic shape of a cubic polynomial after removal of $\nu$ and $\dot\nu$. Therefore, these features are inevitable results of mathematics. However, there is still no self-consistent theory to explain the reason for the sign of $\ddot\nu$. Besides, there exist some residuals in both modes that show more complicated structures beyond the simple mathematics. Therefore, based on the numbers of local extrema, the $W$ mode residuals are further divided into $Wa$, $Wb$, $Wc$ and $Wd$ subclasses corresponding to the residual patterns with two, three, four and more local extreme points (see Fig.\,3a, 3b, 3c and 3d which have the typical characteristic structures in the timing residuals for the different subclasses among the 200 pulsars in the sample of $\sigma_{3}/\sigma_{1} < 0.4$), respectively. Similarly, the $M$ mode residuals are further divided into $Ma$, $Mb$ and $Mc$ subclasses corresponding to the residual patterns with two, three and four local extreme points (see Fig.\,4a, 4b and 4c).

In Table 1 we present the numbers of the different subclasses in our sample. In Figure 6 we plot the $P-\dot P$ diagram for our sample and all the observed pulsars taken from the ATNF Pulsar Catalogue \citep{manchester05}. From the distribution of our sample we see that it covers radio pulsars, millisecond pulsars and high energy pulsars. In the following section, we will attempt to use phenomenological model to reproduce these timing residuals, find the physical process that leads to the phenomena, i.e., timing residuals generally includes positive and negative values of $\ddot\nu$ and some of them have sophisticated structures. Then we will further explain the reason for the
structure of timing residuals of a given pulsar varying with the time span of the data analyzed.

\begin{table}
\centering
\caption{Number of the corresponding types in our sample}
\label{typenum}
\begin{tabular}{c|c|c|c|c}
\hline
\hline
$Type$ &   $Wa$   &   $Wb$    &   $Wc$    &   $Wd$     \\
\hline
$Number$ &   60   &   13    &   22    &   6  \\
\hline
$Type$ &   $Ma$   &   $Mb$    &   $Mc$    \\
\hline
$Number$ &   63   &   19   &   7  \\
\hline
\end{tabular}
\end{table}

\begin{figure}
\centering
\includegraphics[scale=0.6, width=\columnwidth]{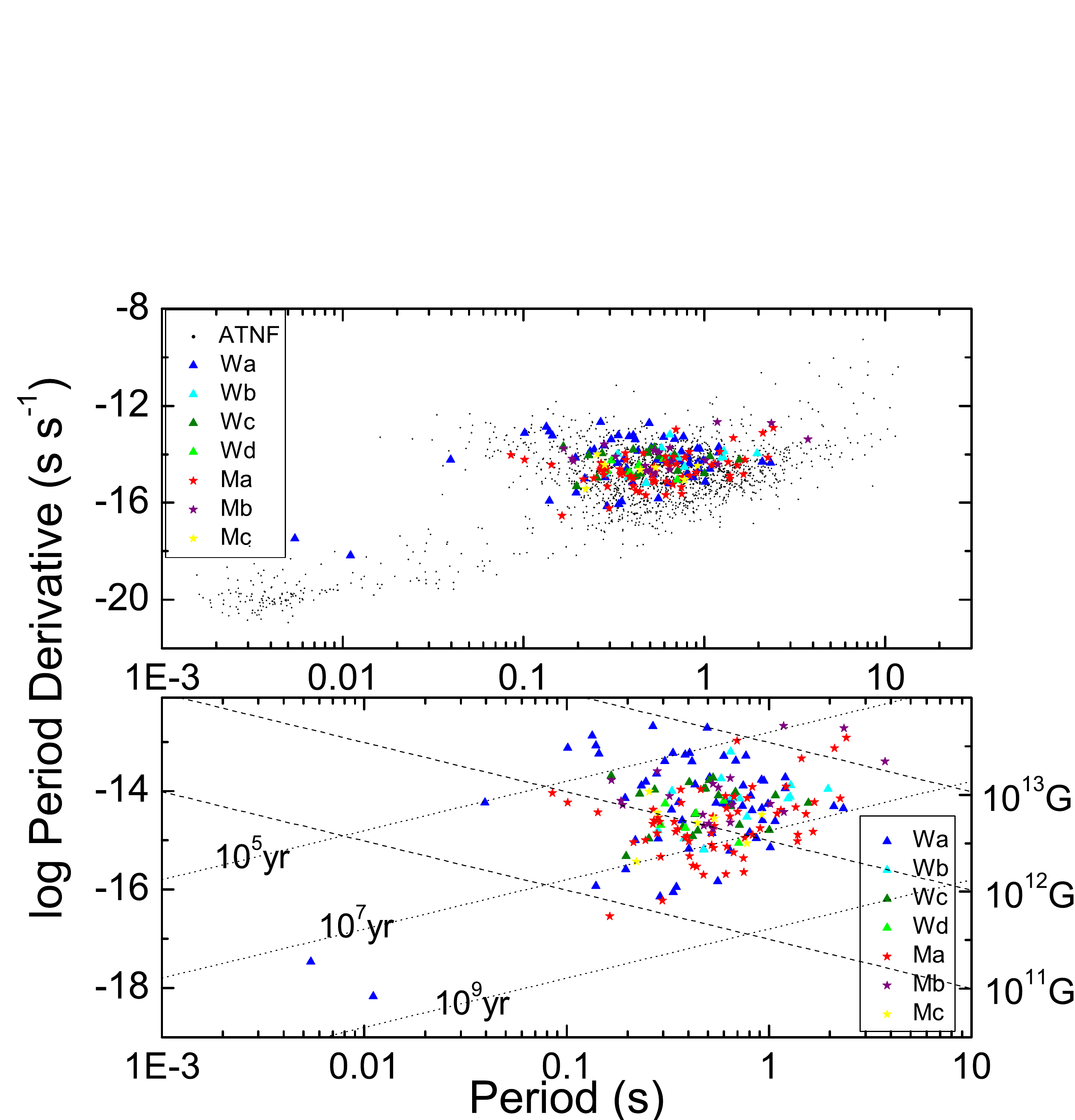}
\caption{The top panel shows the distribution of pulsars in our sample and all other pulsars taken from the ATNF Pulsar Catalog \citep{manchester05} in the $P-\dot P$ diagram. The triangle represents the $W$ mode pulsars, the $M$ mode pulsars are marked with star, and different colors represent the different subclasses. The lower panel is the enlarged view of the distribution of pulsars in our sample; the dotted and dashed lines represent constant characteristic age and constant magnetic field, respectively.}
\label{p-pdot}
\end{figure}

\section{Magnetic Field Decay and Oscillation Model}
\label{sec-model}

For the case of a neutron star in vacuum, due to the asymmetry of magnetic field about its rotation axis, it radiates energy and losses angular momentum. Assuming pure magnetic dipole radiation dominates over the gravitational quadrupole radiation as the braking mechanism for a pulsar's spin down, we have the magnetic-dipole model \citep{pacini67,pacini68, ostriker69}
\begin{equation}\label{dipole}
\dot E_{\rm rot}=I\Omega \dot \Omega=-\frac{2}{3c^{3}}{|\ddot m|^{2}}=-\frac{2(BR^3)^2{sin^2\alpha}}{3c^3}{\Omega ^4},
\end{equation}
or
\begin{equation}\label{other}
\dot\nu = -AB^2\nu^3,
\end{equation}
where $A=\frac{8\pi^{2}R^{6}{\sin^2\alpha}}{3c^3I}$, $m$ is the magnetic dipole moment, $B$ is the strength of the dipole magnetic field at its equatorial surface, $R$ denotes the radius of neutron star, $\alpha$ is the inclination of the magnetic axis with respect to the rotation axis, and $I$ is its moment of inertia. Deriving $\ddot{\nu}$ from equation~(\ref{other}), we have
\begin{equation}\label{nu_ddot1}
\ddot \nu  = 3\dot{\nu}^2/\nu +2\dot{\nu}\dot{B}/B.
\end{equation}
For pulsars with characteristic ages larger than $10^{5}$\,yr in H10 sample, almost equal numbers of pulsars have each sign of $\ddot\nu$. According to equation~(\ref{nu_ddot1}), this phenomenon can be caused by the evolution of magnetic field of second term. However, by considering only monotonic change of magnetic field or torque cannot explain this phenomenon. That is also the reason why other authors suspect that the cubic terms are not due to the intrinsic dipole braking of pulsars \citep{hobbs04}.

However, we have constructed a phenomenological model for the evolution of magnetic field $B$, which contains a long-term power-law decay modulated by short-term oscillations \citep{zhang12a, zhang12b},
\begin{equation}\label{b_decay}
B(t)= B_{\rm L}(t)(1+\sum k_{i}\sin(\phi_{i}+2\pi\frac{t}{T_{i}})),
\end{equation}
where $t$ represents the pulsar's age in arbitrary units, $k_{i}\ll1$, $\phi_{i}$, $T_{i}$ are the amplitude, phase and period of the $i-th$ oscillation component, respectively. Assuming $B_{\rm L}=B_0 ((t+t_{0})/t_{0})^{-\alpha}$, in which $B_0$ is the field strength at age $t_0$, i.e. at the epoch $t=0$, and $\alpha$ is the index of the power law decay. From equation~(\ref{b_decay}), we can obtain the analytic approximation for $\ddot\nu$ (see Zhang $\&$ Xie 2012a Eq. 21),
\begin{equation}\label{ddot_p2}
\ddot{\nu}(t)\simeq 2\dot{\nu}(t)(-\frac{\alpha}{t}+\sum f_i\cos(\phi_i+2\pi\frac{t}{T_i})),
\end{equation}
where $f_i={2\pi k_i/T_i}$, $t$ is the real age of the pulsar. We can see that the second term of equation~(\ref{ddot_p2}) dominates the sign of $\ddot\nu$, if $|\frac{\alpha}{t}|<f_i$. Here we assume that the decay index $\alpha =0.5$ is constant, the value of $k$ and $T$ range from $10^{-5}$ to $10^{-2}$ and from 2 yr to 100 yr, and $f_{\rm max}=1.0\times 10^{-9}$ can be derived \citep{zhang12a}. Therefore, only to the pulsars younger than about 1000 yr, the first term of equation~(\ref{ddot_p2}) can dominate the sign of $\ddot\nu$. Since the ages of the pulsars in our sample are all more than $10^{4}$ yr, the first term of equation~(\ref{ddot_p2}) can be neglected, especially over the short-term evolution. In our phenomenological model, therefore, the sign and magnitude of $\ddot\nu$ can be well described and the structure of timing residuals reflects the oscillation of its magnetic field. Furthermore, for constant $k$ and $T$ (i.e. periodic oscillation), the key factor that influences the sign of $\ddot\nu$ is the variation of the phase of oscillation component.

To further verify our magnetic oscillation model, we try to use it to simulate the observed timing residuals.
Since $(t/t_0)$ and $k$ are small quantities,
\begin{eqnarray}
\centering
B^2(t)&\simeq& B_0^2(1-2\alpha\frac{t}{t_0})(1+\sum2k_i\sin(\phi_i+\omega_i t)) \nonumber\\
&\simeq&B_0^2\left(1-2\alpha\frac{t}{t_0}+\sum2k_i\sin(\phi_i+\omega_i t)\right).\nonumber
\end{eqnarray}

Then, integrating equation~(\ref{other}) and combining with equation~(\ref{b_decay}), we can obtain the expression,
\begin{equation}\label{second}
\frac{1}{\nu^2}=\frac{1}{\nu_0^2}+2AB_0^2\left(t-\frac{\alpha}{t_0}t^2-\sum\frac{2k_i}{\omega_i}\cos(\phi_i+\omega_i t)\right),
\end{equation}
where $\nu_0=\nu(t=t_0)$ and $\omega_i = 2\pi/T_i$.

From equation~(\ref{other}) we can obtain $-AB^2\nu^2=\dot{\nu}/\nu$. $-AB^2\nu^2$ is a small quantity, since $\Delta\nu/\nu\ll1$ during an observation span. Therefore from equation~(\ref{second}), we get,
\begin{equation}\label{third}
\nu=\nu_0-AB_0^2\nu_0^3\left(t-\frac{\alpha}{t_0}t^2-\sum\frac{2k_i}{\omega_i}\cos(\phi_i+\omega_i t)\right).
\end{equation}

Integrating equation~(\ref{third}) gives the expected pulse phase,
\begin{equation}\label{phifan}
\Phi_{m}(t)=\Phi_0+\nu_0t+AB_0^2\nu_0^3(\frac{\alpha}{3t_0}t^3-\frac{t^2}{2}+\sum\frac{2k_i}{\omega_i^2}\sin(\phi_i+\omega_i t)).
\end{equation}

If we define $\dot{\nu}_0\equiv-AB_0^2\nu_0^3$, $\ddot{\nu}_0\equiv2AB_0^2\nu_0^3\alpha/t_0$, $F_i\equiv2k_iAB_0^2\nu_0^3/\omega_i^2$, then equation~(\ref{phifan}) can be rewritten as,
\begin{equation}\label{six}
\Phi_{m}(t)=\Phi_0+\nu_0t+\frac{\dot{\nu}_0t^2}{2}+\frac{\ddot{\nu}_0t^3}{6}+\sum F_i\sin(\phi_i+\omega_i t).
\end{equation}
From equation~(\ref{six}) we can find that the pulse phase can be written as a polynomial to the third order, plus several sinusoidal waves. One thing we would like to emphasize here is that $\ddot{\nu}_0$ we defined in equation~(\ref{six}) is not equivalent to our derived instantaneous $\ddot{\nu}(t)$ in equation~(\ref{ddot_p2}), neither to the averaged $\bar{\ddot{\nu}}$ from conventional fitting observed data span with the third order polynomial of Taylor expansion. The power-law decay of magnetic field always results in $\ddot{\nu}_0>0$, since the decay index $\alpha >0$. In Section 2, the classification of timing residuals of our sample is based on the value of practical fitting averaged $\bar{\ddot{\nu}}$.

In practical procedure of pulsar timing, fitting a polynomial to the third order without sinusoidal waves is a common treatment:

\begin{equation}
\Phi_{\rm{fit}}(t)=\Phi_0+\bar{\nu}t+\frac{\bar{\dot{\nu}}}{2}t^2+\frac{\bar{\ddot{\nu}}}{6}t^3.
\label{taylor}
\end{equation}

By comparing between equations~(\ref{six}) and (\ref{taylor}), we can obtain,
\begin{equation}
\begin{aligned}
\bar{\nu}&=\nu_0+\sum_{i}\xi_{1i}(\phi_{i},\omega_{i}\tau)\omega_{i}F_{i},
\\
\bar{\dot{\nu}}&=\dot{\nu}_0+\sum_{i}\xi_{2i}(\phi_{i},\omega_{i}\tau)\omega_{i}^{2}F_{i},
\\
\bar{\ddot{\nu}}&=\ddot{\nu}_0+\sum_{i}\xi_{3i}(\phi_{i},\omega_{i}\tau)\omega_{i}^{3}F_{i},
\\
\end{aligned}\label{threeterms}
\end{equation}
where $\xi_{1i}$, $\xi_{2i}$ and $\xi_{3i}$ correspond to the coefficients of the first, second and third order terms of polynomial fitting of $\sin(\phi_{i}+\omega_{i}t)$, respectively, which are all functions of time span $\tau$, since:
\begin{equation}
\sin(\phi_{i}+\omega_{i}t)\rightarrow \xi_{1i}(\phi_{i},\omega_{i}\tau)\omega t+\frac{\xi_{2i}(\phi_{i},\omega_{i}\tau)}{2}\omega^{2}t^2+\frac{\xi_{3i}(\phi_{i},\omega_{i}\tau)}{6}\omega^{3}t^3\cdots
\end{equation}
Therefore, from equation~(\ref{threeterms}) we can see that $\bar{\ddot{\nu}}$ is consisted of $\ddot{\nu}_0$ and the third order terms of polynomial fitting of oscillation components. But instantaneous $\ddot{\nu}(t)$ includes the contributions of power-law decay and all the oscillation components at every moment. Only if there is no oscillation in the magnetic field, i.e., $F_i=0$, then $\ddot{\nu}(t)\simeq \bar{\ddot{\nu}}=\ddot{\nu}_0$

\section{Simulations and Fitting Observations}
\label{sec-model}

Therefore, the timing series of phases $\Phi_{s}(t)$ of the simulated TOAs can be obtained based on equation~(\ref{six}). Then, we fit the simulated $\Phi_{s}(t)$ by the second order of Taylor expansion over the observation time span $\tau$:
\begin{equation}\label{Phi}
\Phi_{s}(t)= \Phi_{s0}+\bar{\nu_{0}}(t)+\frac{1}{2}\bar{\dot\nu}t^{2}.
\end{equation}
The simulated timing residuals can be obtained by,
\begin{equation}\label{res}
R_{\rm sim}(t_{i})=\frac{\Phi_{s}(t_{i})-\Phi_{m}(t_{i})}{\nu_0}.
\end{equation}

To extract the best model parameters by comparing with observations, we use {\tt emcee} code from \citep{foreman13}, which is an affine invariant ensemble sampler for MCMC and designed for Bayesian parameter estimation, to sample the full parameter space.
Based on Bayes theorem, assuming $\bmath{p}$ and $\bmath{d}$ represent the model parameters and the data\,(t, residuals, errors.), respectively. Therefore, the posterior probability function can be written as,
\begin{equation}\label{posterior}
  \mathrm{P}(\bmath{p} \, | \, \bmath{d}) \propto
  \mathrm{P}(\bmath{d} \, | \, \bmath{p}) \, \mathrm{P}(\bmath{p}),
\end{equation}
where $\mathrm{P}(\bmath{d} \, | \, \bmath{p})$ is the likelihood function, $\mathrm{P}(\bmath{p})$ is the prior function.

Here we define the likelihood function as a simple Gaussian,
\begin{equation}
    \mathrm{P}(\bmath{d} \, | \, \bmath{p}) = \exp\left(-\frac{\chi^2}{2}\right),
\end{equation}
with
\begin{equation}
    \chi^2 = \sum_{i=1}^{N_{\rm{obs}}} \left(\frac{R_{\rm re}(t_{i})-R_{\rm sim}(t_{i})}{ \sigma_{3} }\right)^2,
\end{equation}
where $N_{\rm{obs}}$ denotes the number of the reported timing residuals, $R_{\rm re}(t_{i})$ and $R_{\rm sim}(t_{i})$ represent the reported timing residuals and the simulated timing residuals at $t_{i}$, respectively. The error of each point of timing residual corresponds to the value of $\sigma_{3}$.

In prior function\,($\mathrm{P}(\bmath{p})$), we need to give physically acceptable ranges of our parameters.
Here we set priors within given bounds:

(i) $\omega_{\rm{min}}<\omega_{i}<\omega_{\rm{max}}$, where $\omega_{\rm{min}}=2.0\times10^{-9}\,\rm{s^{-1}}$, $\omega_{\rm{max}}=1.0\times10^{-7}\,\rm{s^{-1}}$;

(ii) $2k_{\rm{min}}AB_0^2\nu_0^3/\omega_{\rm{max}}^2<F_{i}<2k_{\rm{max}}AB_0^2\nu_0^3/\omega_{\rm{min}}^2$, where $k_{\rm{min}}=10^{-5}$, $k_{\rm{max}}=10^{-2}$ and $\nu_0$ corresponds to the initial value of a pulsar;

(iii) $\phi_{i}$ is between 0 and $2\pi$.

After all these have been prepared, we obtain the best fitting parameters of our model with the following steps.
(1)\,We fit the $R_{\rm sim}(t_{i})$ to the $R_{\rm re}(t_{i})$ preliminarily to find the maximum likelihood. To each parameter, we start by initializing 100 walkers in a tiny Gaussian ball around the maximum likelihood result; (2)\,We sample the parameter space according to the $\mathrm{P}(\bmath{p} \, | \, \bmath{d})$ and regard the set of parameters for which the $\mathrm{P}(\bmath{p} \, | \, \bmath{d})$ is maximized as our best model, since the parameter set can reproduce the $R_{\rm re}(t_{i})$ most closely; (3)\,We can get marginalized distribution for each parameter independently in one-dimension histograms and two dimensional projections of posterior probability distributions\,(In Appendix A, we present the two-dimensional marginalized distributions of all the parameters of the seven typical pulsars); (4)\,We choose the medians of the one-dimension posterior distributions as the best fitting parameters, and the parameter uncertainties are calculated with their 68 percent confidence intervals.

Specifically, the number of parameters in our model is determined by
the number of oscillation parameters of magnetic field. Figures 7 and 8 show the corresponding power spectra of the typical pulsars in Figures 3 and 4, respectively. The number of main peaks in the power spectrum equals the number of oscillation components in our magnetic field evolution model that we assume initially.
Then we give a set of initial values of the oscillation parameters $k_{i}$, $T_{i}$ and $\phi_{i}$ in equation~(\ref{b_decay}) to obtain the timing residuals to be fitted, as well as the initial values of $\nu_0$, $\dot{\nu}_{0}$ and $\ddot{\nu}_0$ and the observation time span $t_s$. The values of $\nu_0$ and $\dot{\nu}_{0}$ are selected from H10 to each pulsar. We keep them constant, since they can hardly change during the short-term evolution. Here we set the power-law index $\alpha =0.5$ and $t_{0}$ as characteristic age to determine the value of $\ddot{\nu}_0$ in equation~(\ref{six}), because the short observation time span can not really constrain the monotonic evolution of the magnetic fields of these pulsars. (We will further discuss if the long-term monotonic evolution of magnetic field would influence the patterns of timing residuals in such short-term observations in section 5.1.) Based on equation~(\ref{res}), to $Wa$, $Wb$, $Ma$ and $Mb$ types, three parameters are needed to be sampled by MCMC\,($F$, $\omega$ and $\phi$). To $Wc$, $Wd$ and $Mc$ types, which include two oscillation component of magnetic field, six parameters should be sampled by MCMC\,($F_{1}$, $\omega_{1}$, $\phi_{1}$, $F_{2}$, $\omega_{2}$ and $\phi_{2}$). After getting the best fitting parameters, we convert them to the parameters we need, i.e., $k_{i}$, $T_{i}$, $\phi_{i}$, and $f_{i}$.

\begin{figure}
\centering
\includegraphics[angle=0,scale=0.4,width=\columnwidth]{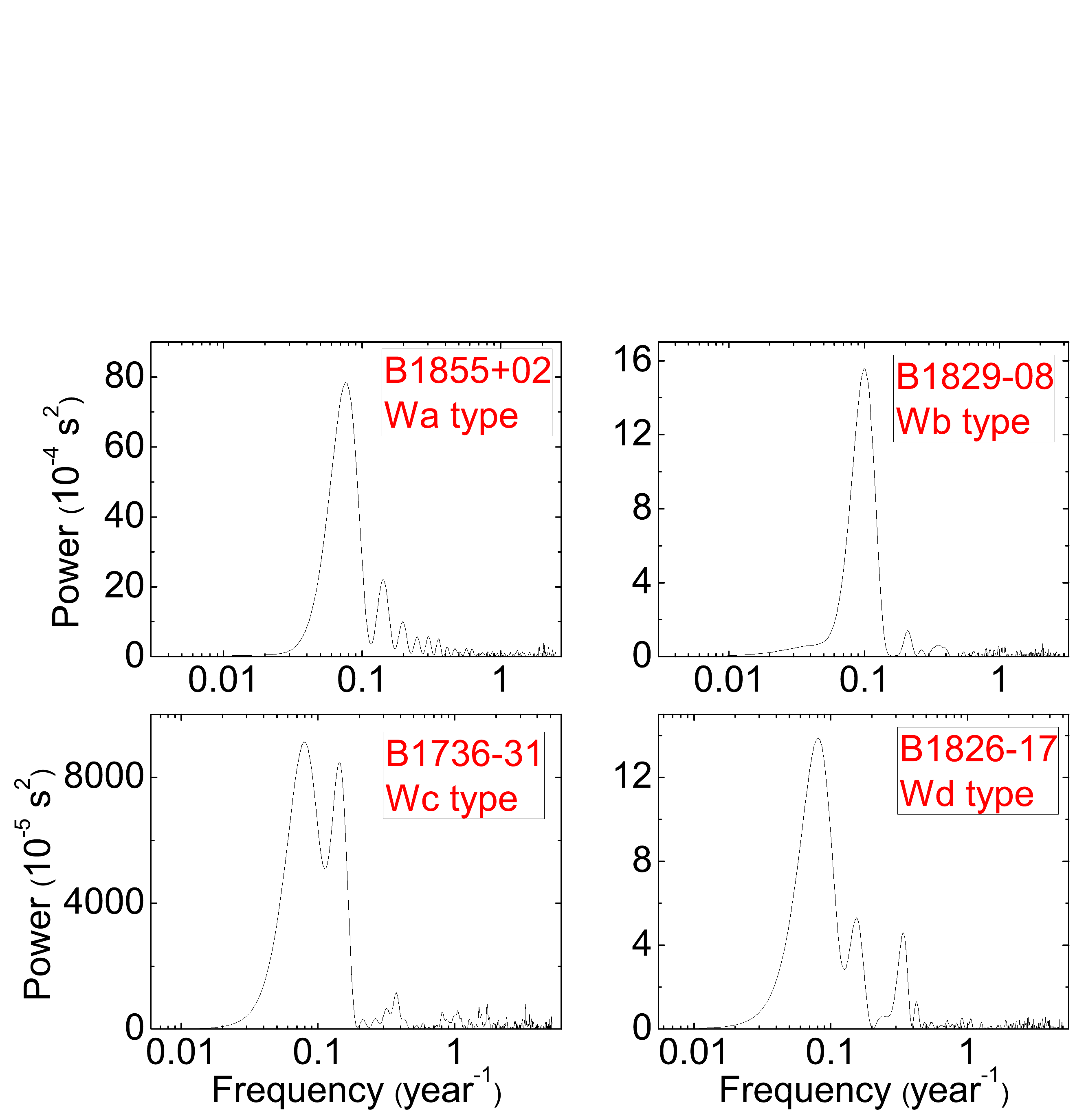}
\caption{Power spectra of B1855+02, B1829-08, B1736-31 and B1826-17.}
\label{W-pwd}
\end{figure}

\begin{figure}
\centering
\includegraphics[angle=0,scale=0.4,width=1\columnwidth]{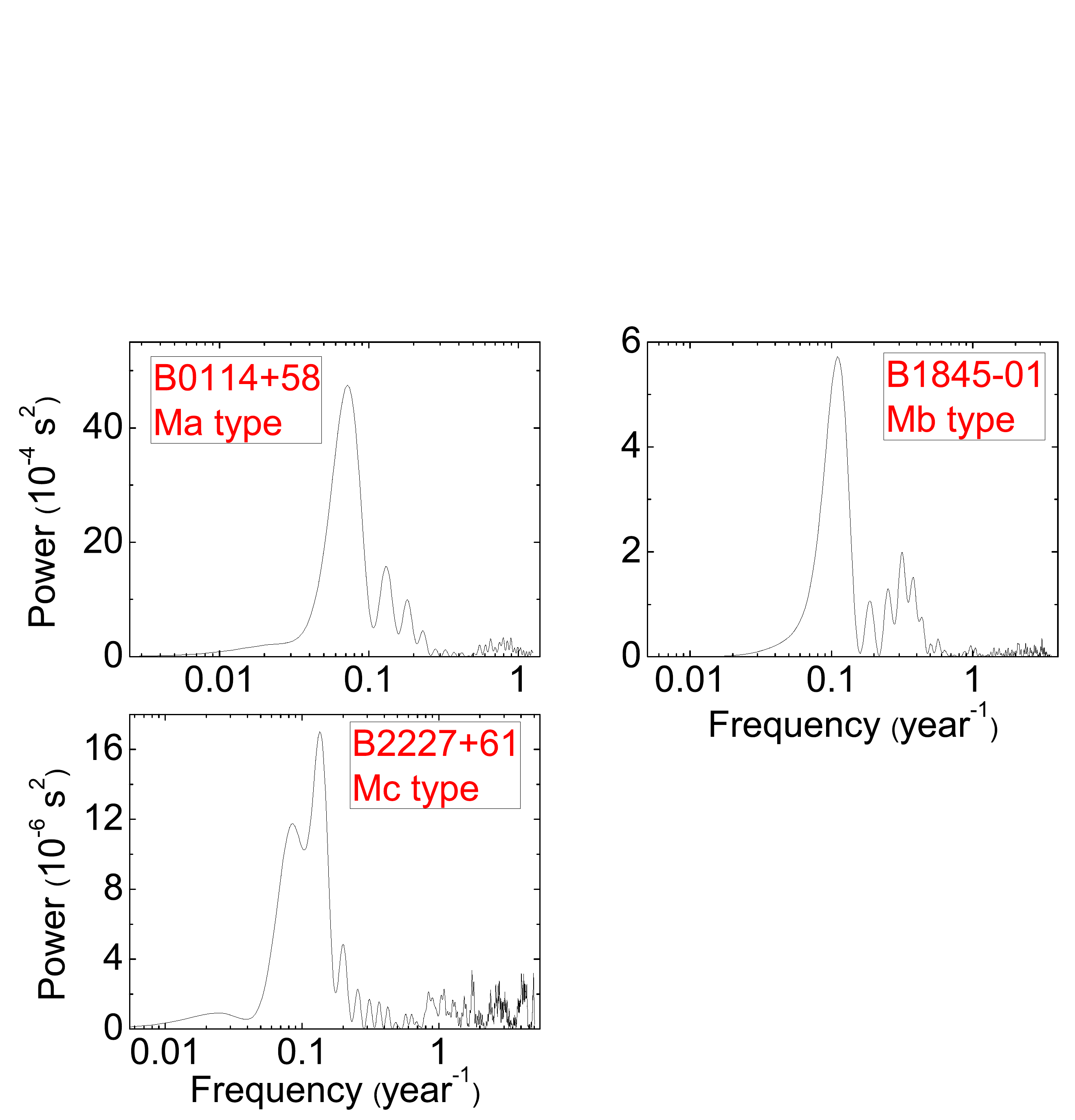}
\caption{Power spectra of B0114+58, B1845-01 and B2227+61.}
\label{M-pwd}
\end{figure}

Figures 9 and 10 show the comparisons between the reported and fitting timing residuals for those several typical pulsars. In the framework of magnetic field oscillation model, we can reproduce almost all types of timing residuals except the $Wd$ type. The $Wd$ type timing residuals of B1826-17 contain three main peaks in its power spectrum; however, we only use two oscillation components to fit the timing residuals to simplify the problem.

\begin{figure}
\centering
\includegraphics[angle=0,scale=0.4,width=\columnwidth]{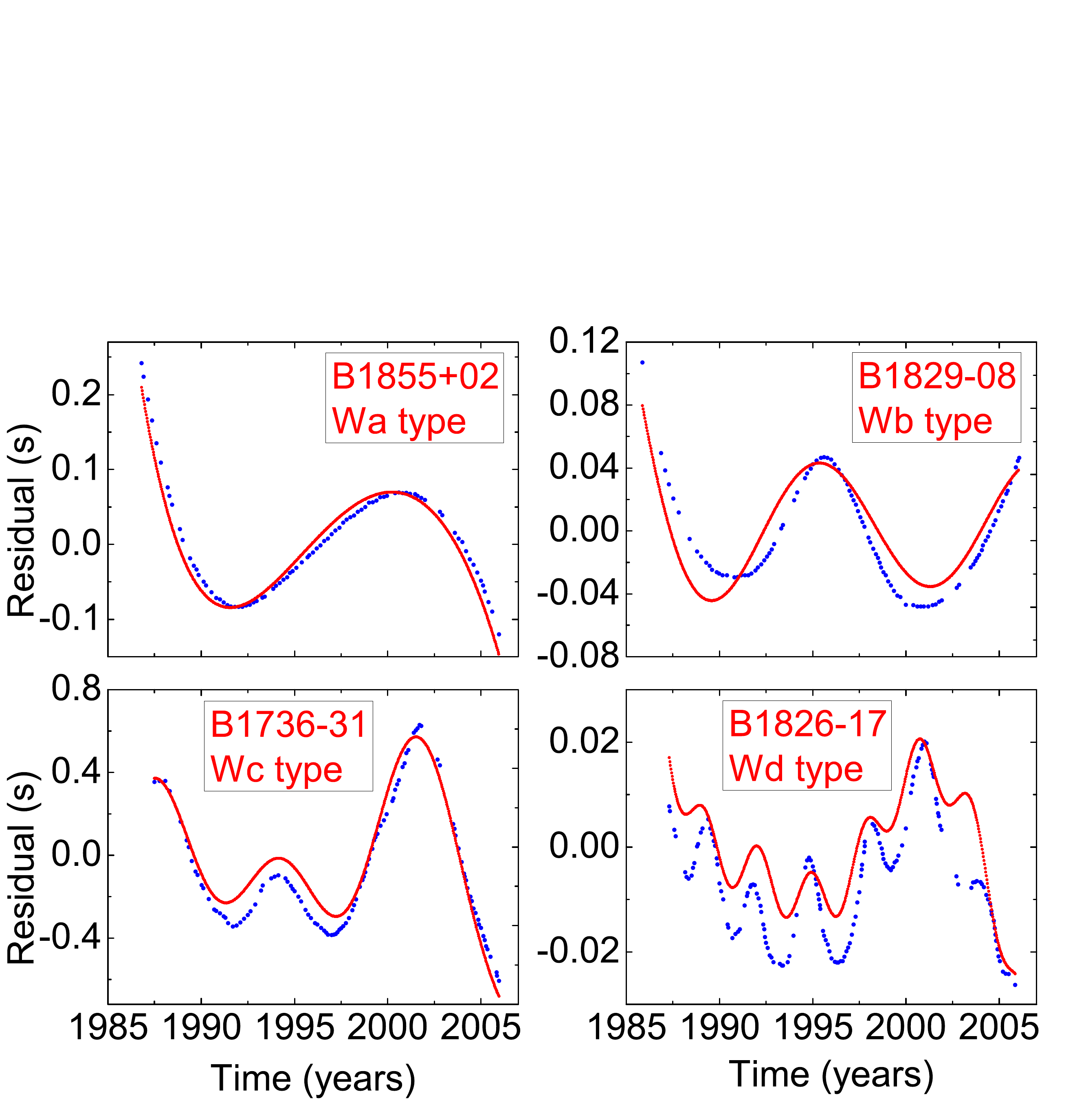}
\caption{Comparisons between our model fitting timing residuals and the corresponding observed timing residuals.}
\label{W-fit}
\end{figure}

\begin{figure}
\centering
\includegraphics[angle=0,scale=0.4,width=\columnwidth]{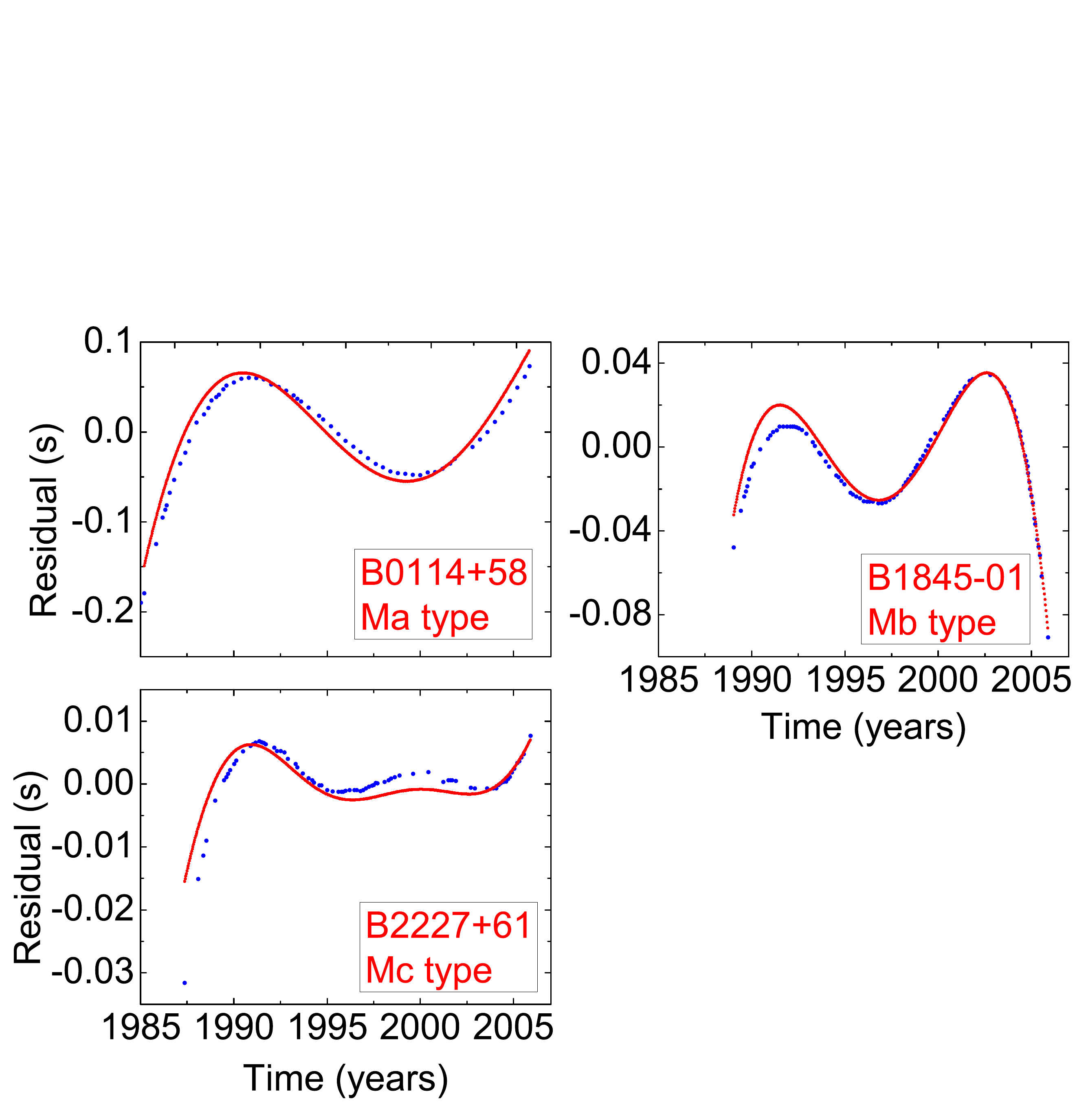}
\caption{Comparisons between our model fitting timing residuals and the corresponding observed timing residuals.}
\label{M-fit}
\end{figure}

In Tables 2 and 3 we present the best fitting parameters which are converted from $F_{i}$, $\omega_{i}$ and $\phi_{i}$, corresponding to the different kinds of timing residuals. Combining equation~(\ref{ddot_p2}) and equation~(\ref{nu_ddot1}), $\sum{f_{i}}\simeq \ddot{\nu}(t)/2\dot{\nu}(t)=\dot{B}/B$, which can reflect the required fractional variation of the magnetic field in the short-term evolution.
Figures 11 and 12 show the comparisons between the power spectra of the fitting timing residuals and the reported timing residuals of typical pulsars.

\begin{table*}
\centering
\caption{Best fitting parameters of the single oscillation component of magnetic field}
\label{parameters}
\begin{tabular}{c|c|c|c|c|c|c}
\hline
\hline
$Name$     & $Type$  & $k\times10^{-4}$ & $T$(yr) & $\phi$ & $f\times10^{-12}$ \\
\hline
B1855+02 & $Wa$ &  $5.83_{-0.18}^{+0.20}$ & $64.48_{-0.43}^{+0.42}$ & ${4.943}_{-0.004}^{+0.004}$ & $1.803_{-0.069}^{+0.072}$ \\
\hline
B1829-08 & $Wb$ &  $0.673_{-0.003}^{+0.004}$ & $13.22_{-0.02}^{+0.02}$ & ${0.379}_{-0.005}^{+0.005}$ & $1.015_{-0.007}^{+0.006}$ \\
\hline
B0114+58 & $Ma$ &  $1.61_{-0.07}^{+0.07}$ & $38.32_{-0.30}^{+0.29}$ & ${5.30}_{-0.02}^{+0.02}$ & $0.84_{-0.04}^{+0.04}$ \\
\hline
B1845-01 & $Mb$ &  $23.9_{-3.5}^{+4.2}$ & $31.93_{-0.85}^{+0.95}$ & ${6.12}_{-0.05}^{+0.05}$ & $14.89_{-2.54}^{+3.10}$ \\
\hline
\end{tabular}
\end{table*}

\begin{table}
\centering
\caption{Best fitting parameters of the double oscillation component of magnetic field}
\label{parameters2}
\begin{tabular}{c|c|c|c}
\hline
\hline
$Name$ & B1736-31 & B1826-17 & B2227+61 \\
\hline
$Type$ & $Wc$ & $Wd$ & $Mc$ \\
\hline
$k_1\times10^{-3}$ & $1.03_{-0.01}^{+0.01}$ & $0.68_{-0.01}^{+0.01}$ & $6.57_{-2.75}^{+4.73}$ \\
\hline
$T_1$(yr) & $16.47_{-0.04}^{+0.04}$ & $2.829_{-0.003}^{+0.003}$ & $64.69_{-6.43}^{+8.50}$ \\
\hline
$\phi_1$ & $5.043_{-0.007}^{+0.006}$ & $0.27_{-0.02}^{+0.02}$ & $5.77_{-0.21}^{+0.23}$ \\
\hline
$f_1\times10^{-11}$ & ${1.249}_{-0.018}^{+0.016}$ & ${4.808}_{-0.075}^{+0.076}$ & ${2.023}_{-0.967}^{+1.841}$ \\
\hline
$k_2\times10^{-5}$ & $263.4_{-0.5}^{+0.3}$ & $9.81_{-0.76}^{+0.82}$ & $179.4_{-90.0}^{+158.1}$ \\
\hline
$T_2$(yr) & $7.210_{-0.002}^{+0.003}$ & $21.70_{-0.32}^{+0.39}$ & $18.98_{-2.10}^{+1.77}$ \\
\hline
$\phi_2$ & $5.076_{-0.002}^{+0.002}$ & $5.86_{-0.03}^{+0.04}$  & $6.19_{-0.40}^{+0.31}$ \\
\hline
$f_2\times10^{-11}$ &  $7.278_{-0.014}^{+0.013}$ & $0.090_{-0.008}^{+0.009}$ & $1.884_{-1.026}^{+2.098}$ \\
\hline
\end{tabular}
\end{table}

\begin{figure}
\centering
\includegraphics[angle=0,scale=0.4,width=\columnwidth]{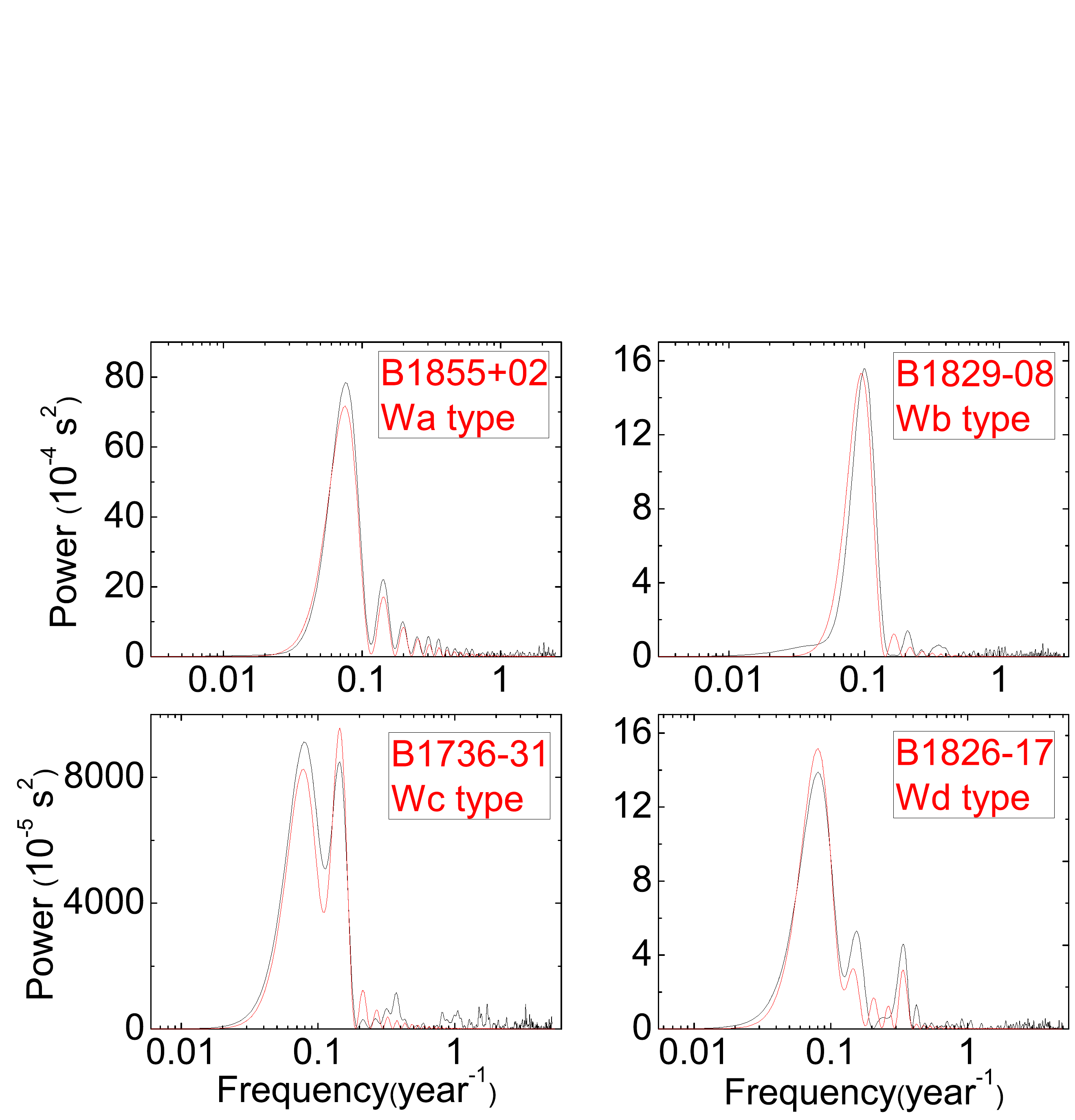}
\caption{Comparisons between the power spectra of our model fitting timing residuals and the corresponding observed timing residuals.}
\label{W-fit-pwd}
\end{figure}

\begin{figure}
\centering
\includegraphics[angle=0,scale=0.4,width=\columnwidth]{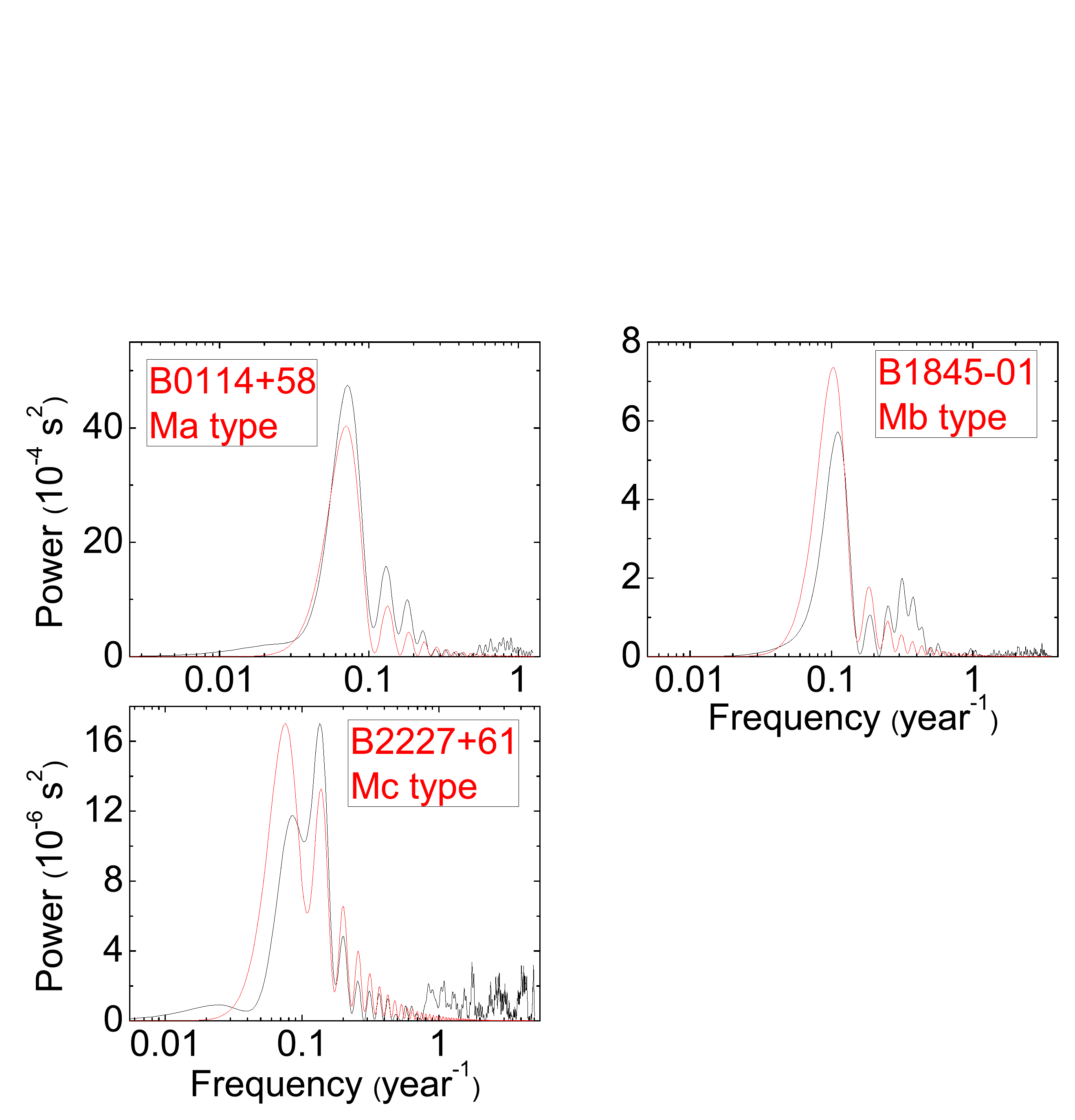}
\caption{Comparisons between the power spectra of our model fitting timing residuals and the corresponding observed timing residuals.}
\label{M-fit-pwd}
\end{figure}

\section{Reclassify the Timing Residuals}
\label{sec-reclassify}

As we pointed out in the previous section, to each oscillation component, the sign of $\bar{\ddot\nu}$ is dominated by the variation of phase, which is an oscillation parameter in our magnetic field oscillation model; namely, the sign of $\bar{\ddot\nu}$ is mainly determined by the observation time. In the meantime, our magnetic oscillation model have reproduced all the typical timing residuals well. Subsequently, we will further attempt to explain the sign of $\bar{\ddot\nu}$ varying with the time span of data analyzed so as to influence the structure of timing residuals. The different types of timing residuals we defined may be caused by the different stages of magnetic field oscillation.

\subsection{Long-term Monotonic Evolution of Magnetic Field}

Here we only discuss two cases of long-term monotonic evolution of magnetic field for simplicity. We first explore if the exponential decay can influence the patterns of timing residuals,
although Zhang \& Xie\,(2012a) have ruled out the exponential decay as the long-term monotonic evolution of the magnetic fields for the sample of H10.
We first assume an exponential field decay $B_{\rm D}{(t)}=B_{\rm 0}e^{-t/\tau_{\rm D}}$, where the timescale of decay $\tau_{\rm D}=10\,\rm {Myr}$ and $B_{\rm 0}$ is the strength of the surface dipole magnetic field at time $t_{\rm 0}=0$, the initial parameters $\bar{\nu}=9.858\,\rm s^{-1}$, $\bar{\dot\nu}=-5.96\times 10^{-13}\,\rm s^{-2}$, and the characteristic age is $2.62\times 10^{5}\,\rm{yr}$; the observation time span we simulate is $\tau=20\,\rm {yr}$. We plot the timing residuals by extracting the first $10$ yr of the simulated observational data segment that is obtained by numerical computation just like in the previous section. Then we extract the data segment from the $1\rm st$ year to the $11\rm th$ year of the simulated observation to plot the timing residuals; the time step is one year and other data segments are taken in a similar fashion, until we extract the data segment from the $11\rm th$ year to the $20\rm th$ year. The upper panel of Figure 13 shows the timing residuals of all the data segments extracted. To the short term observation of the pulsar, we can see that the structures and the magnitude of all the timing residuals are exactly the same during the same observation time interval. The sign and magnitude of $\bar{\ddot\nu}$ do not change, because the observation time span and the structure of timing residuals are all the same.

\begin{figure*}
\centering
\includegraphics[angle=0,scale=0.4,width=0.7\textwidth]{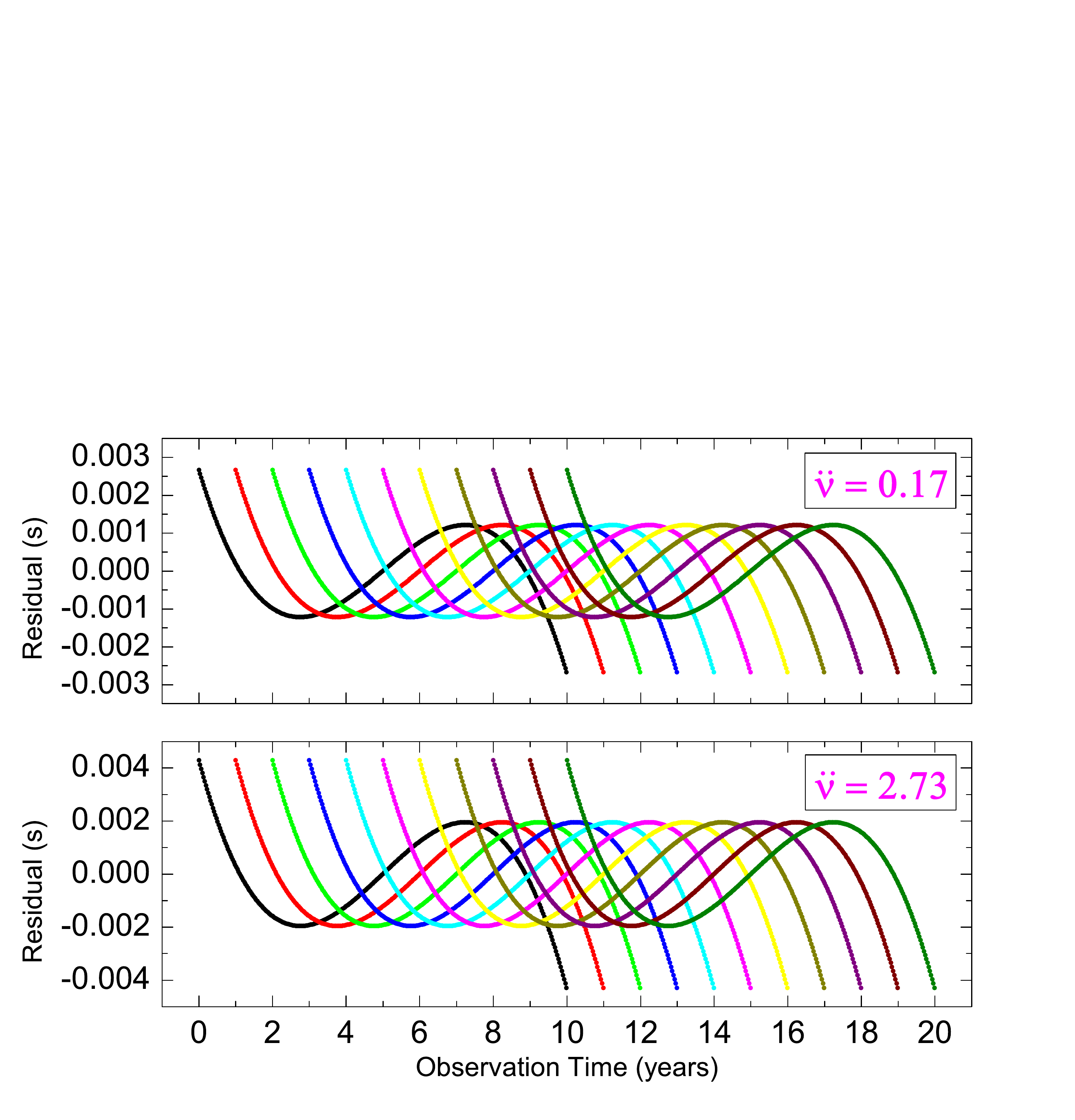}
\caption{Simulated patterns of timing residuals with long-term monotonic decay of magnetic fields of pulsars. The upper panel shows timing residuals simulated with a magnetic field exponential decay model. We assume the initial parameters $\bar{\nu}=9.858\,\rm s^{-1}$ and $\bar{\dot\nu}=-5.96\times 10^{-13}\,\rm s^{-2}$, the observation time span $t_s=20\,\rm {yr}$ and the timescale of decay $\tau_{\rm D}=10\,\rm {Myr}$. The time span of all the timing residuals is 10 yr. $\bar{\ddot\nu}$ is in the units of $10^{-25}\,\rm s^{-3}$.
The lower panel shows timing residuals simulated with a magnetic field power-law decay. The initial parameters and observation time span are kept the same as the conditions of the exponential decay. The decay index $\alpha$ is 0.5. }
\label{edecay}
\end{figure*}

We then assume a power-law field decay $B_{\rm L}{(t)}=B_0 (t+t_{0}/t_{0})^{-\alpha}$, where the decay index $\alpha=0.5$, and $B_{\rm 0}$ is the strength of surface dipole magnetic field at time $t_{\rm 0}=0$. The initial parameters and observation time span are kept the same as the conditions of the exponential decay. The timing residuals are obtained also by the same operation steps as in the previous subsection. The lower panel of Figure 13 shows the timing residuals of all the data segments extracted. The result is nearly the same as for the exponential decay, namely, the structure of timing residuals and the sign of $\bar{\ddot\nu}$ are all independent from the different observation stages in the short term observation. Therefore, the long-term monotonic evolution of magnetic field can not influence the patterns of timing residuals during short-term observations.

\subsection{Periodic Oscillations}

We use our magnetic field oscillation model in equation~(\ref{b_decay}) and adopt the same initial parameters and observation time as before, and assume a single oscillation component, corresponding to the oscillation parameters $k=10^{-4}$, $T=10$ yr and $\phi=0$, respectively. The timing residuals are obtained also
by the same steps above. Figure 14 shows the timing residuals of all the data segments extracted. These timing residuals include $Wa$, $Wb$, $Ma$ and $Mb$ types defined in section 2. The correlation between the sign of $\bar{\ddot\nu}$ and different types of timing residuals is the same as that in observations. From the simulation we also find that, the pulsar keeps $Wa$ and $Ma$ types for most of the time during the periodic oscillation. This result agrees with the fact that $Wa$ and $Ma$ types occupy the majority of all the timing residuals in our sample. This indicates that these four types of timing residuals should be the different oscillation stages of the single oscillation component of magnetic field.

As a consequence, the long-term monotonic decay of magnetic field has no impact to the structure of the timing residuals, but short-term oscillations can explain the observations well. Therefore, the different structures of timing residuals of observations and the corresponding sign of $\bar{\ddot\nu}$ are determined by the observation time, namely, the variation of the oscillation phase $\phi$ of magnetic field. Obviously, our previous classification of timing residuals only according to the tendency of structure does not have physical significance. Hereafter, we reclassify the previously defined timing residuals of $Wa$, $Wb$, $Ma$ and $Mb$ types, which are all reproduced by a single oscillation component of magnetic field, as Single-component Residuals ($SR$) mode; the other types which include double or multiple oscillation components are defined as Multi-component Residuals ($MR$) mode. Since the timing residuals that include double or multiple oscillation components are complex, we do not simulate these situations in this paper.

\begin{figure*}
\centering
\includegraphics[angle=0,scale=0.4,width=0.7\textwidth]{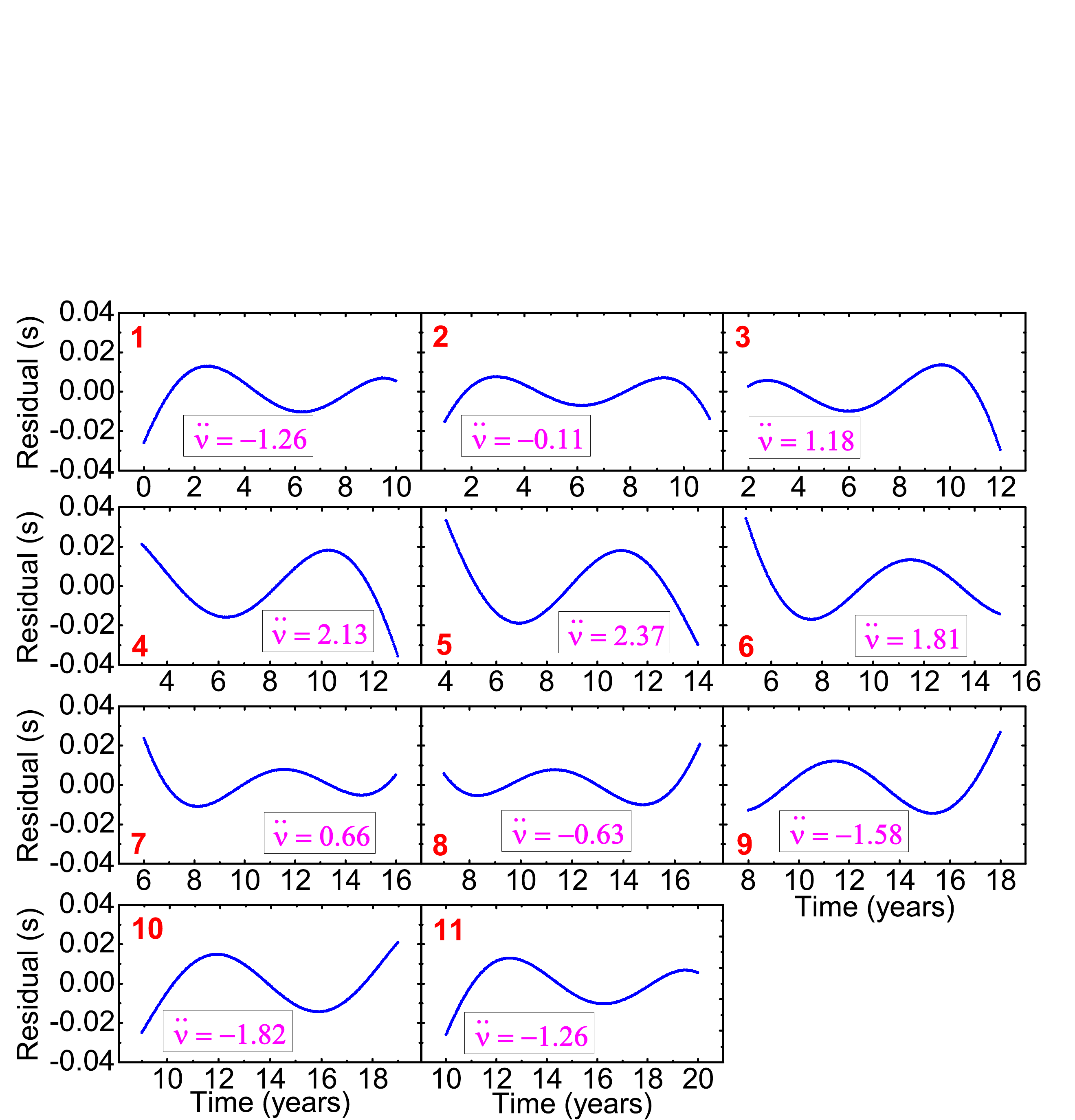}
\caption{Timing residuals simulated with a magnetic field power-law decay and modulated by short-term oscillations model. We assume the initial parameters $\bar{\nu}=9.858\,\rm s^{-1}$ and $\bar{\dot\nu}=-5.96\times 10^{-13}\,\rm s^{-2}$, the observation time span $t_s=20\,\rm {yr}$ and the decay index $\alpha=0.5$. We consider the initial parameters of the single oscillation term, corresponding to $k=10^{-4}$, $T=10\,\rm {yr}$ and $\phi=0$, respectively. The sequence number in every sub-panel represents the chronological order of the extracting data segments. In each sub-panel $\bar{\ddot\nu}$ is in the units of $10^{-25}\,\rm s^{-3}$.}
\label{odecay}
\end{figure*}

\section{Physical Implications of $MR$ mode}
\label{sec-discuss}

Figure 15 shows the distribution of the pulsars of the reclassified timing residuals in the $P-\dot P$ diagram. For the $SR$ mode pulsars, their distribution covers almost all the evolution stages of pulsars, including normal radio pulsars, millisecond pulsars and high energy pulsars. However, the distribution of $MR$ mode pulsars are mainly concentrated in the region of normal radio pulsars.

\begin{figure}
\centering
\includegraphics[angle=0,scale=0.4,width=\columnwidth]{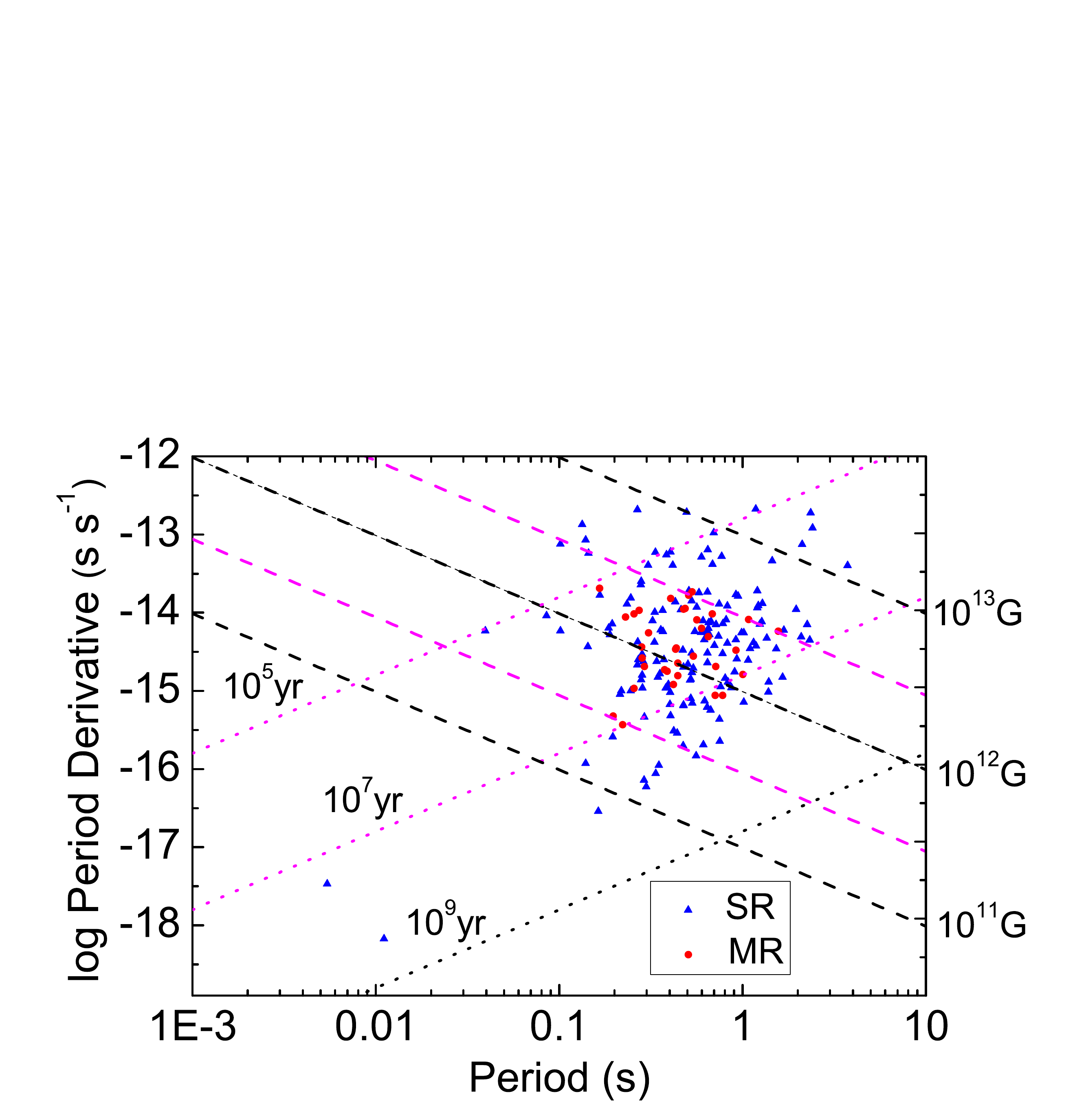}
\caption{The distribution of the reclassified timing residuals in the $P-\dot P$ diagram. $SR$ mode pulsars are marked with triangles, and the circles represent $MR$ mode pulsars. The dashed lines and dotted lines represent constant magnetic field and constant characteristic age, respectively. The Transition Region is confined by the the pink dotted lines and pink dashed lines.}
\label{p-pdot-new}
\end{figure}

To examine if these two samples are derived from the same distribution, we make the two-dimensional Kolmogorov-Smirnov (K-S) test; the returned p-value is 0.13, which means these two samples unlikely originate from the same distribution. Therefore, we speculate that the physical processes that show a single oscillation form exist over the whole evolution of pulsars all along. However, as the neutron stars evolve to normal pulsars, there may exist a transition region ($TR$), where double or multiple oscillation components could appear in some pulsars. Afterwards the single oscillation mode dominates again, accompanied with the evolution of pulsars. $TR$ is shown in Figure 15 with the pink dashed lines and dotted lines, where the magnetic field is from $3\times 10^{11}\ \rm G$ to $3\times 10^{12}\ \rm G$ and the characteristic age is from $10^{5}$ yr to $10^{7}$ yr. The $TR$ phenomenon may be indirect evidence of long term magnetic field evolution of pulsars.

In both isolated and accreting neutron stars, the Hall effect and Ohmic decay are both very important to the long term magnetic field evolution \citep{cumming14}. Since the Hall drift conserves magnetic energy, it cannot be a direct cause of magnetic field decay. However, Jones (1988) first demonstrated that Hall drift could transport magnetic field from the inner crust to the outer crust, where Ohmic dissipation is much more effective than in the inner crust. Hall drift can also pump energy from an internal strong toroidal field to the poloidal component, resulting in increase of magnetic field \citep{pons12, gourgouliatos14}. It indicates that Hall drift may produce damped oscillation with long period between toroidal and poloidal field components \citep{marchant14}. Cumming et al. (2004) proposed that the Hall effect dominates over ohmic decay after the time $t_{\rm switch}\simeq10^{4}B_{12}^{-3}$ yr, when the magnetization parameter, $R_{B}=\frac{t_{\rm Ohm}}{t_{\rm Hall}}>1$, where $t_{\rm Ohm}$ and $t_{\rm Hall}$ represent the timescales of Ohmic decay and Hall effect, respectively. The Hall timescale at lower densities of the crust is \citep{cumming14}
\begin{equation}\label{Hallout}
\tau_{\rm Hall,~Outer}=\frac{5.7\times 10^4~{\rm
yr}}{B_{12}}\rho_{12}^{5/3}(\frac{Y_e}{0.25})^{11/3}(\frac{g_{14}}{2.45})^{-2},
\end{equation}
where $\rho_{12}=\rho/10^{12}~{\rm g~cm^{-3}}$, $Y_e$ is the number fraction of electrons, and $g_{14}$ is the local gravity, assumed constant.

Gourgouliatos $\&$ Cumming (2014) investigated the evolution due to Hall drift using different initial conditions, and found that eventually the field evolves towards a similar configuration defined as `attractor', which consists of a dominant dipolar poloidal field coupled to a octopole through a weak toroidal quadrupole. Recently Marchant et al. (2014) did a similar numerical simulation about the effect of Hall drift to magnetic field evolution in neutron stars. The initial condition $E_{p}/E=0.9$ represents the ratio of poloidal to total energy, namely the poloidal component contains a large amount of energy, and $R_{B}=100$ means that Hall drift dominates the evolution. From the result of the numerical simulation, we find that the current associated to the toroidal field drags the poloidal field lines closer to one of the poles, since the bending of the poloidal field lines changes the orientation of the toroidal field; therefore, the poloidal field lines are dragged to the opposite pole. This stable oscillation is repeated until the evolution is dominated by Ohmic dissipation (see Fig.\,1 in Marchant et al 2014). From Fig. 1 in Marchant et al. (2014) we estimate the beginning of the oscillation is nearly $\tau_{\rm Hall}$; after approximately 100 $\tau_{\rm Hall}$, the oscillation vanishes when it reaches the attractor state. According to equation~(\ref{Hallout}), we estimate that the evolution time is from $0.6\times 10^{5}$ yr to $0.6\times 10^{7}$ yr, consistent with the $TR$ defined above. It is reasonable to speculate that this stage of oscillations between the poloidal and toroidal components is the main stage to produce the $MR$ mode timing residuals. We thus further argue that the evolution of the modes of timing residuals may have a relation with the long term evolution of magnetic field configurations.

\section{Summary and Conclusion}
\label{sec-summary}

In this paper, we separate the pulsars that have more detailed structure in the timing residuals than others from the H10 sample as our study sample ($\sigma_{3}/\sigma_{1} < 0.4$). According to the sign of $\bar{\ddot\nu}$, we have classified the timing residuals as $W$ mode and $M$ mode, which are further subdivided into several kinds of subclasses based on the structure of timing residuals (Fig.\,3 and Fig.\,4). Then we used our phenomenological model to fit the timing residuals of the typical pulsars by the MCMC method respectively (Fig.\,9 and Fig.\,10) and obtained the best fitting oscillation parameters to all the types of timing residuals. After that, we simulated the short term observation of an individual pulsar with different magnetic field evolution models. Furthermore, we reclassified the timing residuals as $SR$ mode and $MR$ mode based on the number of oscillation components of magnetic field. We also analyzed the physical implications from the distribution of timing residuals in the $P-\dot P$ diagram and the physical mechanisms causing the observational behaviors of timing residuals. Our main results and conclusions are summarized as follows.

1. Our magnetic field oscillation model can reproduce the general behaviors of timing residuals well.

2. The variation of the observed $\bar{\ddot\nu}$ and the detailed structure of timing residuals reflect the oscillation of the magnetic field of pulsars.

3. We rule out long-term monotonic evolution of magnetic field as the underlying mechanism for the observational timing residual patterns.

4. The different structures of timing residuals of observation and the corresponding sign of $\bar{\ddot\nu}$ are determined by the observation time, namely, the variation of oscillation phase $\phi$ of magnetic field.

5. A single magnetic field oscillation mode exists generally over the whole evolution of pulsars, while $MR$ mode only exists in the region of normal radio pulsars.

6. The $SR$ mode and $MR$ mode may transit from each other with the evolution stage of magnetic field configurations.

In this work we only fit the timing residual patterns of seven typical pulsars in the sample of 200 pulsars with obvious timing residual patterns, in order to understand the underlying mechanism of the observational residual patterns and their variations with observation time. In our next work, we will fit the data for all other pulsars to understand the statistical properties of the parameters in our model.

If we can understand better about the behavior of timing residuals, we may be able to improve the sensitivity of using pulsars to detect gravitational waves and explore the physical processes in the interiors of neutron stars. We suggest that our simulation results could be confirmed by using the existing data and predict that the structure of timing residuals should vary along with different segments of data. In doing so, we could further understand the behavior of magnetic field oscillation of pulsars.

\section*{Acknowledgements}

SNZ acknowledges partial funding support by 973 Program of China under grant 2014CB845802, by the National Natural Science Foundation of China (NSFC) under grant Nos. 11133002 and 11373036, by the Qianren start-up grant 292012312D1117210, and by the Strategic Priority Research Program ``The Emergence of Cosmological Structures'' of the Chinese Academy of Sciences (CAS) under grant No. XDB09000000. JNF acknowledges partial funding support by the Joint Fund of Astronomy of NSFC and CAS through grant U1231202, and 973 Program under grants 2014CB845700 and 2013CB834900.

\appendix
\section{2D marginalized distributions of all the parameters}\label{sec:2D}

We obtain two dimensional marginalized distributions of all the parameters of our model with the following steps.

(1)\,To sample the  posterior distribution of parameters, we use {\tt emcee} code from \citep{foreman13}, which provides a fast and stable implementation of an affine-invariant ensemble sampler for MCMC. We can get the $R_{\rm sim}(t_{i})$ through the analytic expression of pulse phases\,(see equation~(\ref{six}) and equation~(\ref{res})).

(2)\,We fit the $R_{\rm sim}(t_{i})$ to the $R_{\rm re}(t_{i})$ preliminarily to find the maximum likelihood. To each parameter, we start by initializing 100 walkers in a tiny Gaussian ball around the maximum likelihood result.

(3)\,We sample the parameter space according to the $\mathrm{P}(\bmath{p} \, | \, \bmath{d})$
regarding the set of parameters for which the $\mathrm{P}(\bmath{p} \, | \, \bmath{d})$ is maximized as our best model, since the parameter set can reproduce the $R_{\rm re}(t_{i})$ most closely. In this process, we also need to output the time series of each parameter in the chain, which shows the parameter values for each walker at each step in the corresponding chain. The walkers start in small distributions around the maximum likelihood values and then they quickly wander and start to explore the full posterior distribution. Therefore, we can check if all the parameters are convergent through this.

\begin{figure*}
\center
\includegraphics[width=0.8\textwidth]{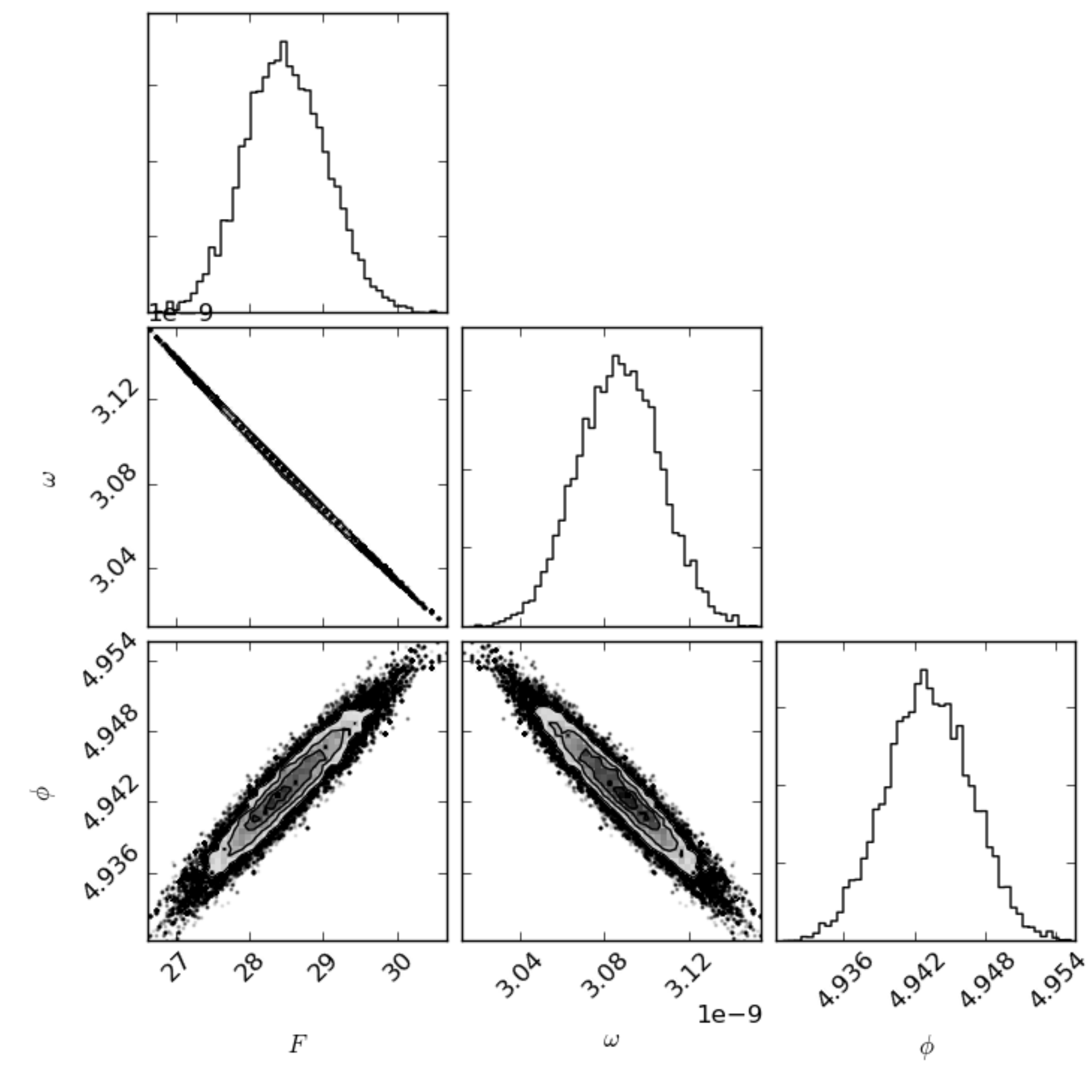}
\caption{This figure shows all the one and two dimensional projections of the posterior probability distributions of parameters of $Wa$ type pulsar B1855+02. Along the diagonal shows the marginalized distribution for each parameter independently in the histograms and the marginalized two dimensional distributions are in the other panels.
\label{Wa-corner}}
\end{figure*}

\begin{figure*}
\center
\includegraphics[width=0.8\textwidth]{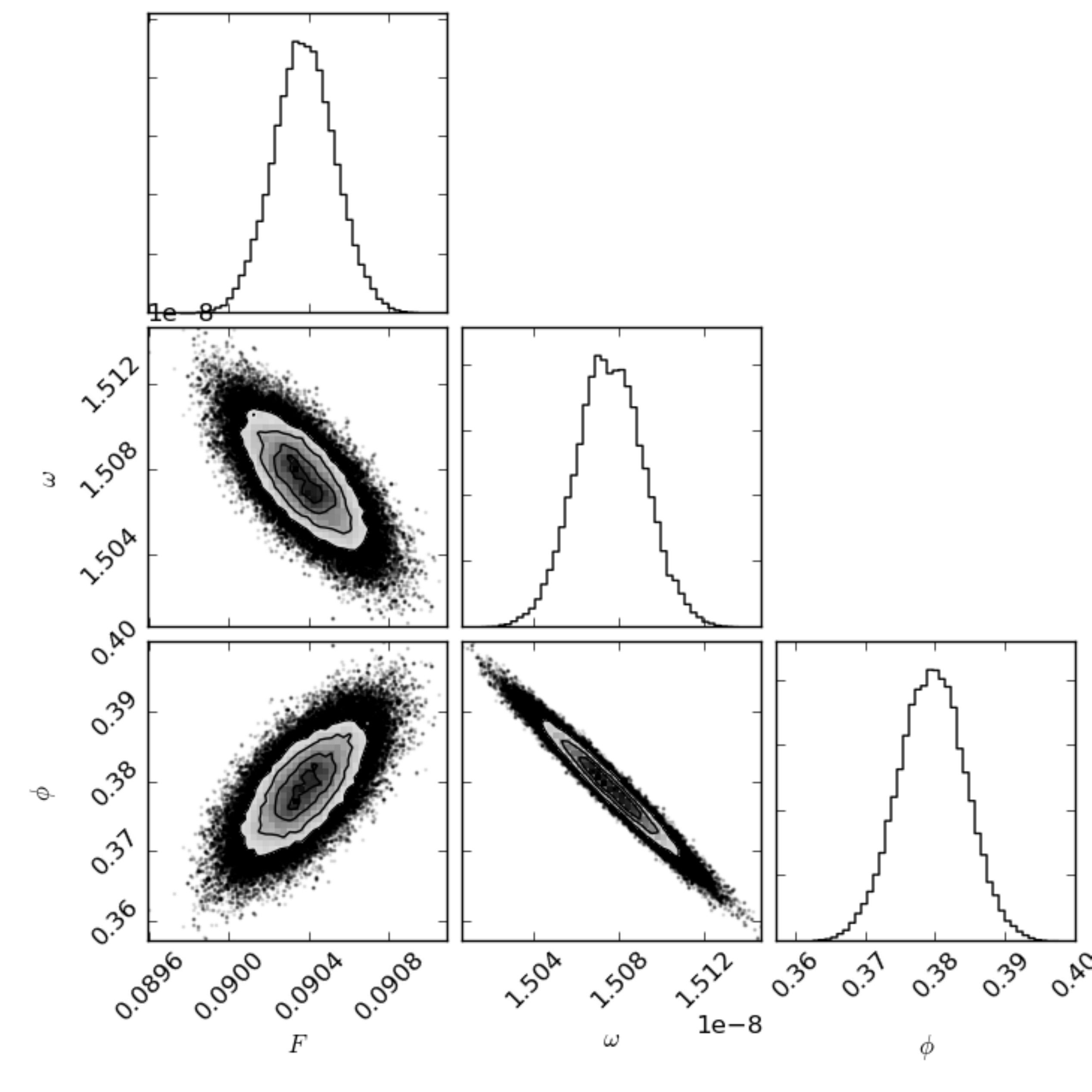}
\caption{Same as Fig.\,\ref{Wa-corner}, but for Wb type pulsar B1829-08.
\label{Wb-corner}}
\end{figure*}

\begin{figure*}
\centering
\includegraphics[width=0.8\textwidth]{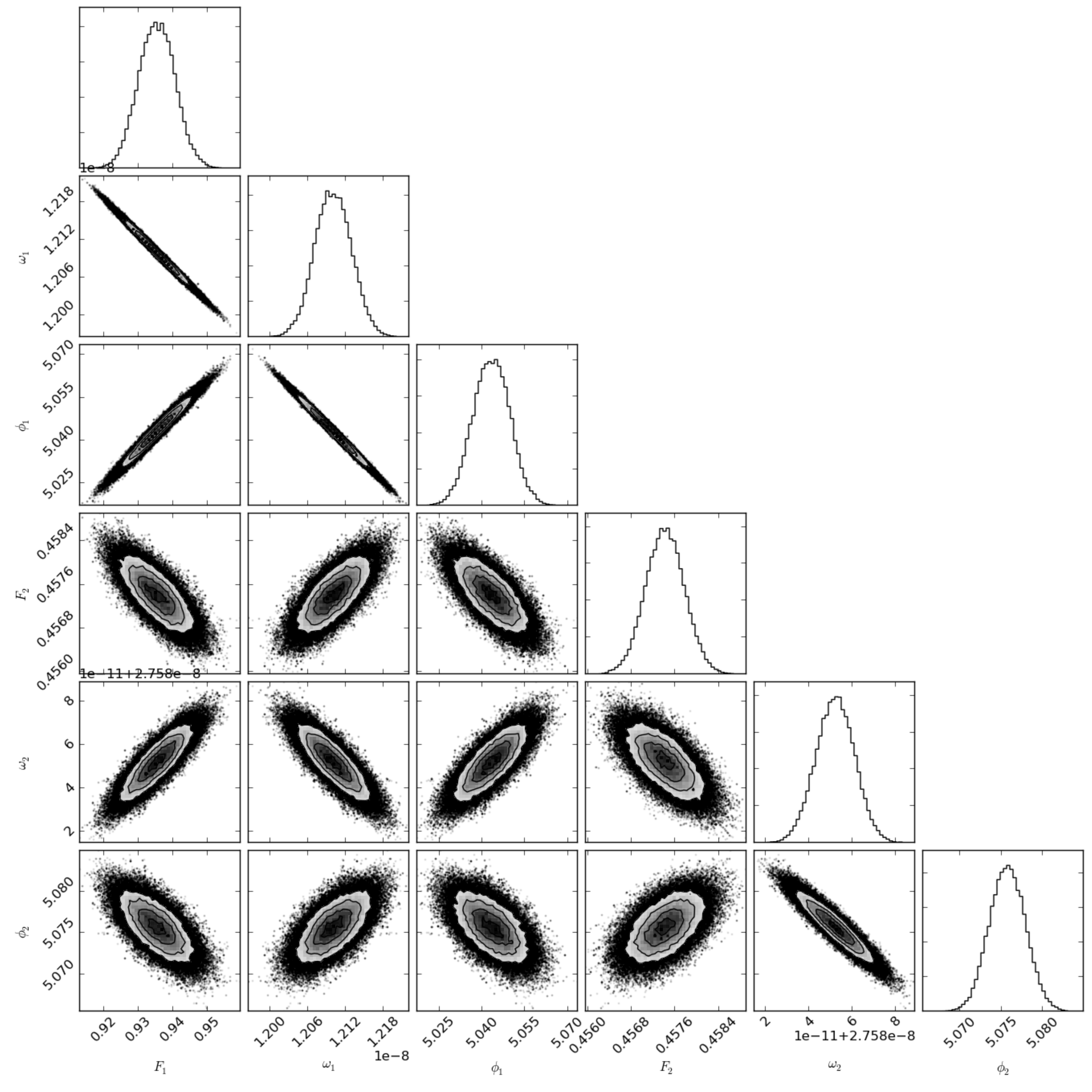}
\caption{Same as Fig.\,\ref{Wa-corner}, but for Wc type pulsar B1736-31.
\label{Wc-corner}}
\end{figure*}

\begin{figure*}
\centering
\includegraphics[width=0.8\textwidth]{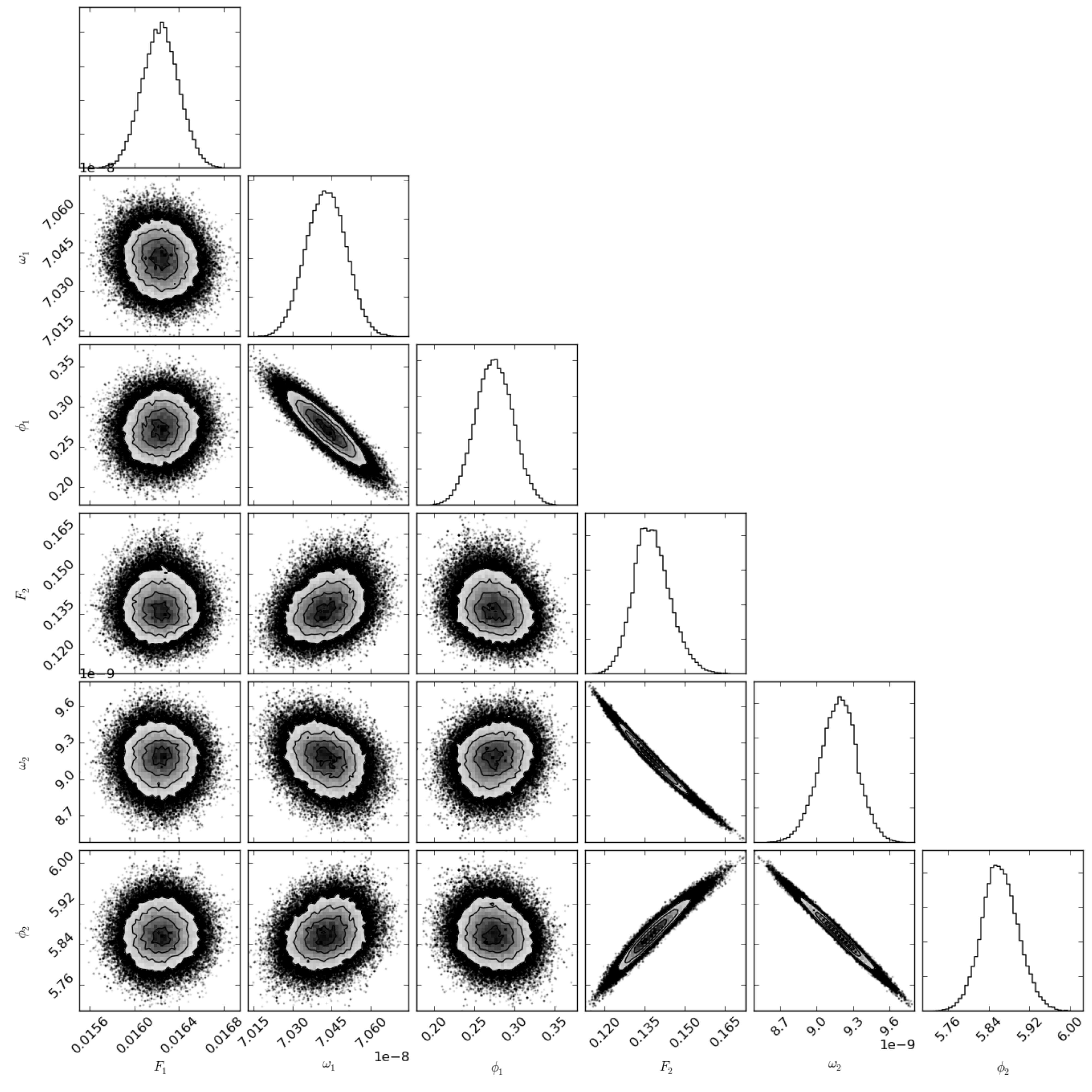}
\caption{Same as Fig.\,\ref{Wa-corner}, but for Wd type B1826-17.
\label{Wd-corner}}
\end{figure*}

\begin{figure*}
\centering
\includegraphics[width=0.8\textwidth]{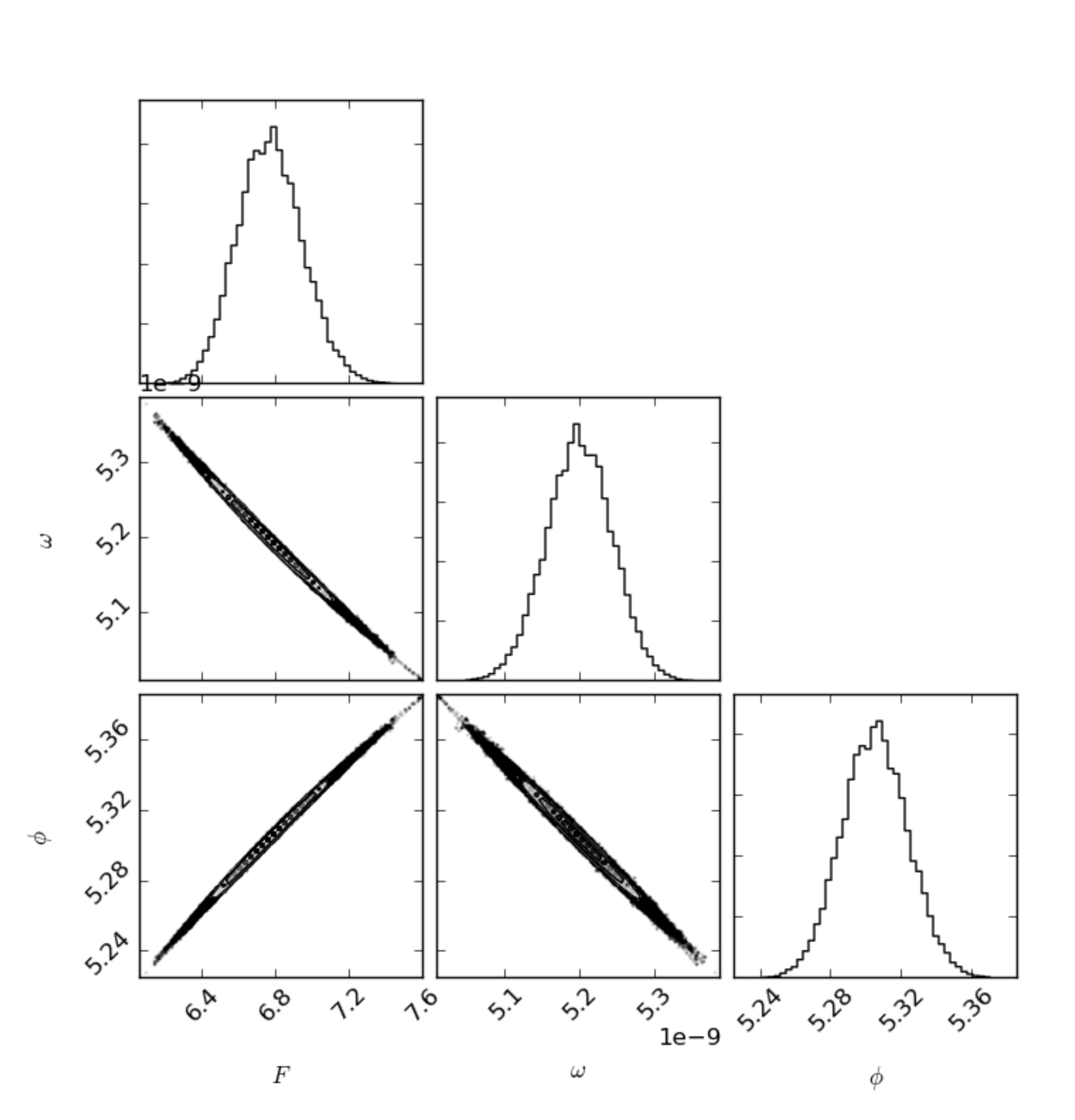}
\caption{Same as Fig.\,\ref{Wa-corner}, but for Ma type pulsar B0114+58.
\label{Ma-corner}}
\end{figure*}

\begin{figure*}
\centering
\includegraphics[width=0.8\textwidth]{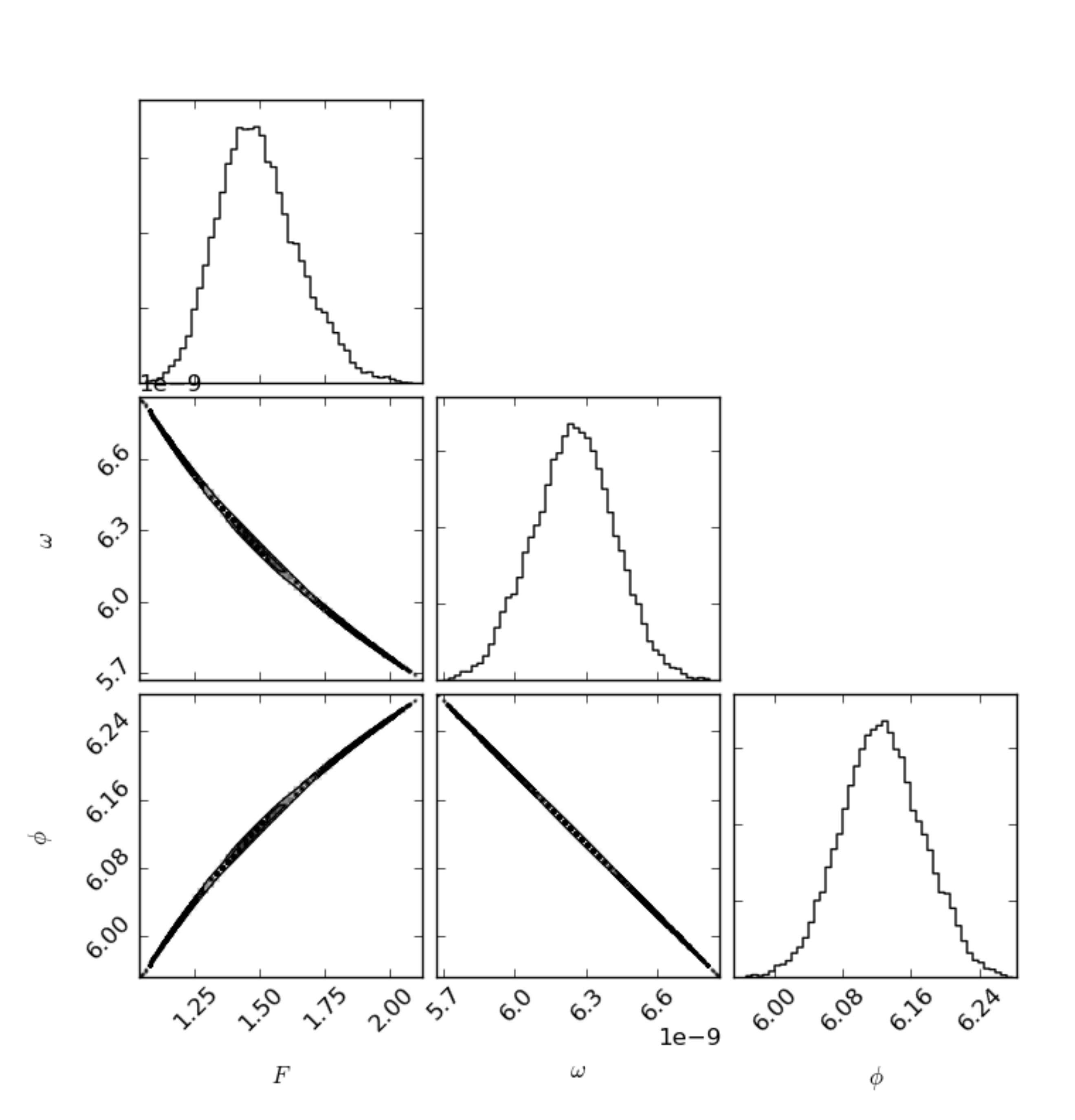}
\caption{Same as Fig.\,\ref{Wa-corner}, but for Mb type pulsar B1845-01.
\label{Mb-corner}}
\end{figure*}

\begin{figure*}
\centering
\includegraphics[width=0.8\textwidth]{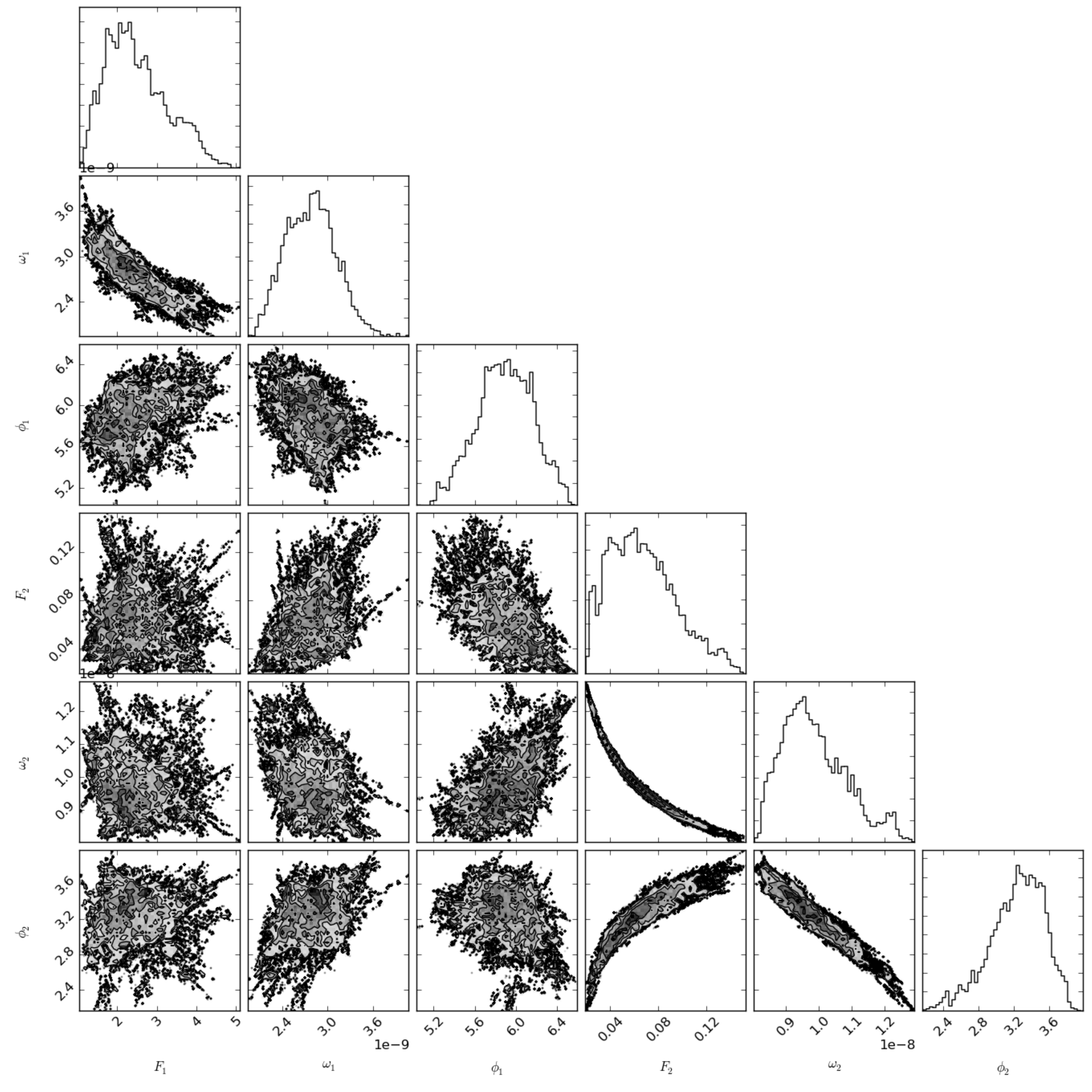}
\caption{Same as Fig.\,\ref{Wa-corner}, but for Mc type pulsar B2227+61.}
\label{Mc-corner}
\end{figure*}

(4)\,After making sure all the parameters are convergent, we can get marginalized distribution for each parameter independently in one-dimension histograms and two dimensional projections of posterior probability distributions\,(see Fig.\,\ref{Wa-corner}, Fig.\,\ref{Wb-corner}, Fig.\,\ref{Wc-corner}, Fig.\,\ref{Wd-corner}, Fig.\,\ref{Ma-corner}, Fig.\,\ref{Mb-corner} and Fig.\,\ref{Mc-corner}).

(5)\,We choose the medians of the one-dimension posterior distributions as the best fitting parameters, and the parameter uncertainties are calculated with their 68 percent confidence intervals.

(6)\,We need to make the projection of our result into the corresponding observed timing residuals.

To further make sure that our best fitting parameters are credible, after getting the final posterior distribution, we also chose 500 samples from the time series of parameters in the chain randomly to produce 500 simulated timing residuals. The result shows that they all overlap well with the `best' simulated timing residuals, which are produced by our best fitting parameters.

\end{document}